\documentclass[11pt]{article}

\usepackage{siunitx}
\usepackage{amsmath}
\usepackage{graphicx}
\usepackage{natbib}

\usepackage[paperwidth=7.87in,paperheight=10.5in,
            textwidth=39pc,textheight=9in]{geometry}

\DeclareMathOperator{\N}{\mathit{N}}
\newcommand{\Np}[2]{\N \left( #1, #2 \right)}

\newcommand{\btheta  }{\boldsymbol{\theta  }}
    
\title{State-space modeling of intra-seasonal persistence in daily climate indices: a data-driven approach for seasonal forecasting}

\title{State-space modeling of intra-seasonal persistence in daily climate indices: a data-driven approach for seasonal forecasting}

\author{Philip G. Sansom, David B. Stephenson and Daniel B. Williamson \\ University of Exeter, Exeter, United Kingdom}

\date{}

\begin{document}

\maketitle

\abstract{
Existing methods for diagnosing predictability in climate indices often make a number of unjustified assumptions about the climate system that can lead to misleading conclusions.
We present a flexible family of state-space models capable of separating the effects of external forcing on inter-annual time scales, from long-term trends and decadal variability, short term weather noise, observational errors and changes in autocorrelation.
Standard potential predictability models only estimate the fraction of the total variance in the index attributable to external forcing.
In addition, our methodology allows us to partition individual seasonal means into forced, slow, fast and error components.
Changes in the predictable signal within the season can also be estimated.
The model can also be used in forecast mode to assess both intra- and inter-seasonal predictability.
\paragraph*{•}
We apply the proposed methodology to a North Atlantic Oscillation index for the years 1948--2017.
Around \SI{60}{\percent} of the inter-annual variance in the December-January-February mean North Atlantic Oscillation is attributable to external forcing, and \SI{8}{\percent} to trends on longer time-scales.
In some years, the external forcing remains relatively constant throughout the winter season, in others it changes during the season.
Skillful statistical forecasts of the December-January-February mean North Atlantic Oscillation are possible from the end of November onward and predictability extends into March.
Statistical forecasts of the December-January-February mean achieve a correlation with the observations of 0.48.
}

\section{Introduction}
\label{sec:intro}

Many processes in the climate system are only active or only interact  intermittently.
Snow cover changes the interaction between the land surface and the atmosphere \citep[e.g.,][]{Chapin2010}.
Sea ice alters the coupling between the ocean and the atmosphere \citep[e.g.,][]{Bourassa2013}.
Stratosphere-troposphere coupling are known to influence storm tracks and surface weather \citep[e.g.,][]{Kidston2015}.
El Ni\~{n}o and La Ni\~{n}a events influence weather patterns in remote regions \citep[e.g.,][]{Toniazzo2006}.
Intermittent forcing is often associated with unusually persistent weather conditions \citep{Bronnimann2007,Tomassini2012}.
Therefore, such events are a potential source of skill for short-term climate forecasts \citep{Sigmond2013,Scaife2014}.

The concept of potential predictability in climate indices was introduced by \citep{Madden1976} and formalized by \citet{Zwiers1987} using analysis of variance methods.
The observed variance in the climate index is partitioned into inter-annual and intra-seasonal components.
Observations within seasons are used to estimate the inter-annual variance in the seasonal mean index that might be expected due to unforced natural variability, i.e., weather.
This estimate is compared with the observed inter-annual variance of the seasonal mean index to determine if some part of the observed process might be attributed to external forcing and be ``potentially predictable''.

Existing methods for assessing potential predictability have two main limitations.
First, the climate is assumed to be stationary (constant mean and variance) throughout the study period.
Fixed trends and seasonal cycles are often estimated and removed from the data.
Gradual changes in the mean or seasonal cycle due to natural or anthropogenic forcing can be confounded with potentially predictable signals on seasonal time-scales, artificially inflating the signal or making it hard to detect.
Second, data are often split into arbitrarily defined seasons and analyzed separately.
Analyzing seasons separately leads to a loss of information by ignoring components common to all seasons.
Conversely, if a predictable signal only exists for part of a season, then it may be difficult to detect over whole season.
In addition, separate autocorrelation functions are often estimated for each season.
This makes it difficult to distinguish changes in the inter-annual variance of the seasonal mean, from changes in the autocorrelation structure.

\cite{Sansom2018a} proposed a more flexible state-space time series approach to assessing potential predictability that directly addresses the limitations of existing methods.
The new methodology allows a whole time series to be analyzed without being split into seasons. 
The timing of the external forcing can be inferred from the data.
Predictable behavior due to temporary changes in the mean can be distinguished from behavior due to temporary changes in the autocorrelation structure.
Non-stationary mean, trend and seasonal components are estimated simultaneously to avoid confounding other potentially predictable signals.
Gradual changes in the autocorrelation structure can also be estimated.

In addition to estimating the fraction of inter-annual variance explained by external forcing, the new methodology allows us to address a number of more detailed questions.
For example, was a particular extreme season the result of external forcing, or simply natural variability?
To answer this, individual seasonal means can be decomposed into mean, weather, error and externally forced components.
Using the new methodology, the external forcing effect can vary continuously, rather than being fixed throughout a season.
Therefore, we can examine changes in the forcing within seasons, and persistence between seasons.
The new statistical model can also be used to forecast climate indices, either by rapidly learning the state of the external forcing within a season, or by exploiting inter-annual persistence in the forced component.

\citet{Sansom2018a} demonstrated their methodology on a short time series of a daily North Atlantic Oscillation (NAO) index and used a complex and expensive Markov Chain Monte Carlo (MCMC) procedure to estimate key parameters.
In this study, we develop a simplified maximum likelihood approach to parameter estimation that is more efficient and can be easily implemented without specialist statistical knowledge.
We analyze a longer 70-year daily NAO index computed from a different data source and compare the results to previous potential predictability studies.
Results are presented for all seasons, rather than just boreal winter and summer.
The longer time series allows us to investigate multi-decadal trends in the NAO and time-varying skill in seasonal forecasts.
We also compare the skill of seasonal forecasts from the statistical model with that of the state-of-the-art GloSea5 seasonal forecasting system \citep{Maclachlan2015}.
Our code is freely available to enable our results to be reproduced, and our methodology to be easily applied to other climate indices.

The remainder of this study is structured as follows, Section~\ref{sec:data} describes the construction and features of the NAO index.
In Section~\ref{sec:methodology} we summarize the statistical model proposed by \citet{Sansom2018a} for diagnosing persistence and predictability.
Section~\ref{sec:modeling} outlines the process of model building, parameter estimation by maximum likelihood, and model checking.
In Section~\ref{sec:results} we present the results of our analysis of the the 70-year NAO index.
Section~\ref{sec:discussion} concludes with a discussion.

\section{Data}
\label{sec:data}

In this study, we analyze a daily NAO index computed from the NCEP/NCAR reanalysis \citep{Kalnay1996}.
Following \citet{Stephenson2006}, the NAO index is calculated as the difference in area averaged mean sea level pressure between two rectangular regions stretching from \ang{20}--\ang{55}N and \ang{55}--\ang{90}N, both spanning \ang{90}W--\ang{60}E.
This definition provides a consistent daily index for the whole year that is robust to seasonal changes in the centers of action of the NAO.
By using a single definition of the NAO throughout the year we avoid possible inconsistencies where multiple definitions are joined together.
Working directly with a pressure index rather than a principle component index improves the interpretability of the model.
By working in the natural units of pressure, we can more easily incorporate quantitative judgments about model components and assess whether our inferences are intuitively reasonable.

\begin{figure*}[t]
  \centerline{\includegraphics[width=\textwidth]{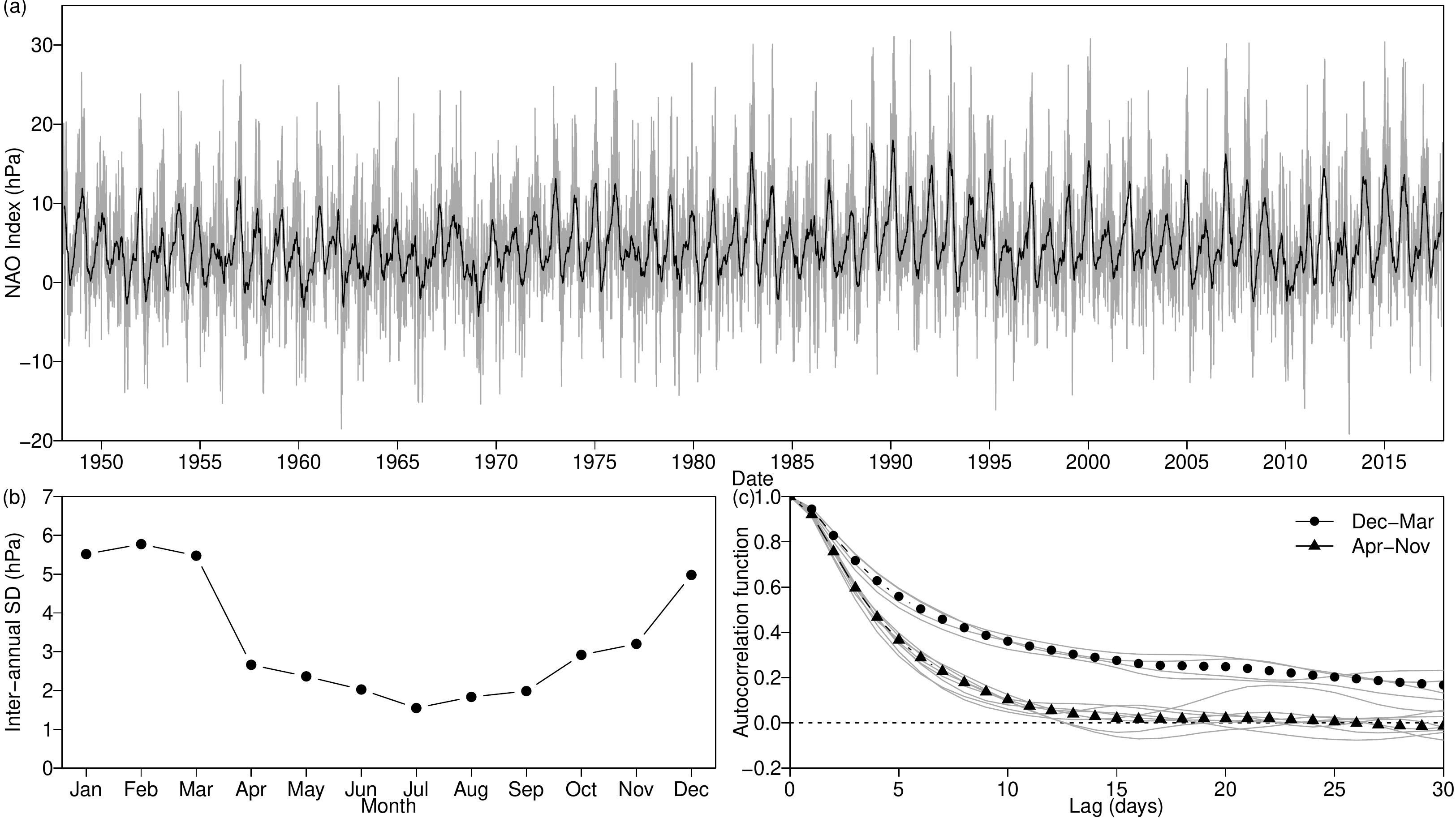}}
  \caption{The North Atlantic Oscillation index.
           (a) Daily time series (grey) and 90 day moving average (black) of our NAO index,
           (b) the inter-annual standard deviation of the monthly mean NAO index, and
           (c) the autocorrelation function of the daily NAO index computed for each month of the year.}
  \label{fig:nao}
\end{figure*}

The daily NAO index computed from the NCEP/NCAR reanalysis is shown in Figure~\ref{fig:nao}(a) and spans the period 1 January 1948 to 31 December 2017, a total of 25,568 daily values.
An annual cycle is clearly visible.
Figure~\ref{fig:nao}(a) shows the inter-annual variance of the monthly mean NAO index.
The inter-annual variance during December--March is between two and three times higher than during April--November.
Figure~\ref{fig:nao}(b) shows the autocorrelation function of the daily NAO index for each month of the year, after detrending and deseasonalizing the data.
The autocorrelation in the NAO index is also stronger during December--March than during April--November.
Figures~\ref{fig:nao}(b) and (c) suggest a persistent change in the dynamics of the daily NAO index during December--March.

\section{Methodology}
\label{sec:methodology}

\subsection{Potential predictability}

The concept of potential predictability for climate indices was formalized by \citet{Zwiers1987} (see also Chapter 17.2 of \citet{vonStorchZwiers}).
One of the simplest formulations of the basic statistical model for potential predictability is
\begin{align*}
  Y_{i,j} & = Z_j + X_{i,j}             
    &  Z_j & \sim \Np{0}{\sigma_Z^2}  \notag \\
  X_{i,j} & = \phi X_{i-1,j} + w_{i,j}  
    &  w_{i,j} & \sim \Np{0}{\sigma_w^2}
\end{align*}
where $Y$ is the observed climate index and the index $i$ denotes the time within a season and $j$ denotes the year.
The time series $X_{i,j}$ represents noise due to weather and is modeled as an autoregressive process with coefficient $\phi$ ($-1 < \phi < 1$), forced by a normal random walk (Gaussian white noise) $w_{i,j}$.
The quantity $Z_j$ represents the externally forced signal and is assumed independent and identically distributed between years $j$.
The underlying assumption is that $Z_j$ varies slowly (i.e., predictably) and can be assumed constant within a particular season, while $X_{i,j}$ varies rapidly throughout the season.
The total variance in $Y_j$ is partitioned into inter-annual variance due to $Z_j$ and intra-seasonal variance due to weather $X_{i,j}$.

The basic potential predictability model above implicitly assumes that the long-term mean of the observed process $Y$ is zero.
If this is not the case, then the mean must first be estimated and removed.
In order to estimate potential predictability in different seasons, the model must be fitted to each season separately.
Fitting each season separately makes it difficult to distinguish changes in the variance of the seasonal mean $\sigma_Z^2$ from changes in the day-to-day persistence $\phi$.

\subsection{A more flexible approach}
\label{sec:model}

\citet{Sansom2018a} proposed a more flexible statistical model for diagnosing persistence and predictability in time series, which we summarize here.
A climate index index $Y_t$ ($t = 1,\ldots,T$) is modeled as the sum of mean, seasonal, weather and intermittently forced components
\begin{align*}
  & \textbf{Forecast equation} \notag \\
  Y_t              & = \mu_t + \textstyle\sum_{k = 1}^K \psi_{k t} + X_t + Z_t + v_t &
    v_t                & \sim \Np{0}{V} \\
  & \textbf{Mean component} \notag \\
  \mu_t            & = \mu_{t-1} + \beta_t + w_{\mu t}  &
    w_{\mu t}          & \sim \Np{0}{W_\mu} \\ 
  \beta_t          & = \beta_{t-1} + w_{\beta t}  &
    w_{\beta t}        & \sim \Np{0}{W_\beta} \\
  & \textbf{Seasonal component} \notag \\
  \psi_{k t}       & = \psi_{k,t-1}       \cos k \omega 
                     + \psi_{k,t-1}^\star \sin k \omega
                     + w_{\psi_k t} &
    w_{\psi_k t}       & \sim \Np{0}{W_\psi} \\
  \psi_{k t}^\star & = \psi_{k,t-1}^\star \cos k \omega   
                     - \psi_{k,t-1}       \sin k \omega
                     + w_{\psi_k^\star t} &
    w_{\psi_k^\star t} & \sim \Np{0}{W_\psi} \\
  & \textbf{Weather component} \notag \\
  X_t              & = \textstyle\sum_{p=1}^P \phi_{p t} X_{t-p} + w_{X t} &
    w_{X t}            & \sim \Np{0}{W_{X t}}  \\
  \phi_{p t}       & = \phi_{p,t-1} + w_{\phi_p t} &
    w_{\phi_p t}       & \sim \Np{0}{W_\phi} 
\end{align*}
for $k = 1,\ldots,K$ and $p = 1,\ldots,P$ where $\omega = 2 \pi / 365.25$.
The residual $v_t$ represents observation or measurement error.
The expected size of the errors is controlled by the variance $V$.
The parameters $\mu_t$ and $\beta_t$ represent the mean and trend respectively.
The harmonic parameters $\psi_{k t}$ and $\psi_{k t}^\star$ ($k = 1,\dots,K$) represent seasonal or other cyclic behavior.
Gradual changes in the mean, trend and seasonal parameters are captured by the normal random walks $w_{\mu t}$, $w_{\beta t}$, $w_{\psi_k t}$ and $w_{\psi_k^\star t}$.
The variances $W_\mu$, $W_\beta$ and $W_\psi$ control the rate of change of each component.
The weather component $X_t$ represents the day-to-day variability in the NAO index and is modeled as a \emph{time-varying} autoregressive process \citep{PradoWest}.
Gradual changes in the daily autocorrelation structure are captured by normal random walks $w_{\phi_p t}$ on the autoregressive coefficients $\phi_{1 t},\ldots,\phi_{P t}$.
The variance $W_\phi$ controls the rate of change of the autoregressive coefficients.
For many climate indices, including the NAO, the day-to-day variance $W_{X t}$ of the weather component $X_t$ will vary with the annual cycle and can be modeled as
\begin{align*}
  W_{X t} & = W_X + \sqrt{a^2 + b^2}+ a \sin \omega t + b \cos \omega t &
    W_X > 0.
\end{align*}

\subsubsection*{The intermittently forced component}

\begin{figure}[t]
  \centerline{\includegraphics[width=0.5\textwidth]{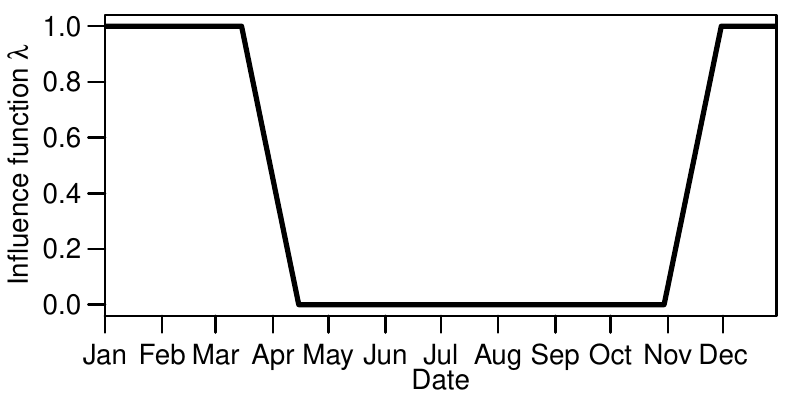}}
  \caption{The influence function $\lambda_t$ chosen for the NAO index.
           The intermittent forcing only affects the observed index when $\lambda_t > 0$. }
  \label{fig:lambda}
\end{figure}

We consider two alternative models for the effect of intermittent forcing leading to unusual persistence -- a temporary change in the mean, or a temporary change in the day-to-day persistence of the weather conditions.
Figure~\ref{fig:nao} suggests a change in the persistence of the NAO between December and March.
We represent the timing of the timing of the forcing effect using an indicator variable
\begin{align*}
  \lambda_t = 
    \begin{cases}
      0 & \text{in Apr--Nov} \\
      1 & \text{in Dec--Mar}.
    \end{cases}
\end{align*}
In practice, we allow a brief transition period between the two regimes by interpolating between 0 and 1 over a short time.
We will refer to $\lambda_t$ as the influence function and the period when $\lambda > 0$ as the forcing period.
Figure~\ref{fig:lambda} shows the influence function chosen for the NAO index in Section~4

A temporary shift in the mean can be captured by introducing a new variable $\delta_t$ and writing the intermittently forced component as
\begin{align*}
  Z_t = \lambda_t \delta_t
\end{align*}
where the shift $\delta_t$ is modeled as
\begin{align*}
  \delta_t & = \varphi \delta_{t-1} + w_{\delta t} &
    w_{\delta t} & \sim \Np{0}{W_\delta}.
\end{align*}
The mean shift $\delta_t$ is analogous to the seasonal mean $Z_j$ in the basic potential predictability model, but varies continuously rather than being assumed constant throughout a season and independent between years.
The shift only affects the observed process $Y_t$ when $\lambda_t > 0$.
Depending on the values of $\varphi$ ($-1 < \varphi < 1$) and $W_\delta$ ($W_\delta > 0$) the mean shift $\delta_t$ can persist between winters ($\varphi \approx 1$, $W_\delta \approx 0$) or vary during a single season ($\varphi < 1$, $W_\delta > 0$).

A temporary change in the day-to-day persistence can be represented by writing the intermittently forced component as
\begin{align*}
  Z_t = \lambda_t \sum_{p=1}^{P} \delta_{p t} X_{t-p}.
\end{align*}
where the $P$ new variables $\delta_{p t}$ alter the autocorrelation structure of the weather component $X_t$ and are modeled as
\begin{align*}
  \delta_{p t} & = \varphi \delta_{p,t-1} + w_{\delta_p t} &
    w_{\delta_p t} & \sim \Np{0}{W_\delta}.
\end{align*}
If the winter behavior in Figure~\ref{fig:nao} is caused by a change in the persistence of day-to-day weather conditions, then we might expected the change to be similar each year, i.e., $\varphi \approx 1$ and $W_\delta \approx 0$.

\subsection{Model fitting}

\citet{Sansom2018a} developed efficient model fitting procedures based on the Extended Kalman Filter \citep[Chapter 10]{DurbinKoopman}.
Model outputs take the form of plausible trajectories for the time evolution of each component $\mu_t$, $\beta_t$, $\psi_{1 t},\psi_{1 t}^\star,\ldots,\psi_{K t},\psi_{K t}^\star$, $X_{t}$, $\phi_{1 t},\ldots,\phi_{P t}$ and $\delta_t$.
Both point and interval estimates can be constructed from samples of plausible trajectories for any combination of parameters or time points of interest.
Forecasting from the fitted model can be achieved simply by simulating from the equations in the previous section.
See \citet{Sansom2018a} for full details.

\section{Modeling the North Atlantic Oscillation index}
\label{sec:modeling}

In order to fit the model described in the previous section to our NAO index, we need to specify the following quantities:
\begin{itemize}
  \item the number of harmonic components $K$;
  \item the order of the autoregressive process $P$;
  \item the variances $V$, $W_\mu$, $W_\beta$, $W_\psi$, $W_{X t}$ and $W_\phi$;
  \item the forcing parameters $\varphi$ and $W_\delta$;
  \item the influence function $\lambda_t$;
  \item initial guesses for $\mu_t$, $\beta_t$, $\psi_{1 t},\psi_{1 t}^\star,\ldots,\psi_{K t},\psi_{K t}^\star$, $X_{t},\ldots,X_{t-P+1}$, $\phi_{1 t},\ldots,\phi_{P t}$ and $\delta_t$ at time $t=0$.
\end{itemize}

Exploratory analysis of the daily NAO index suggests a model with annual and semi-annual cycles ($K=2$) for the seasonal component, and a fifth order autoregressive model ($P=5$) for the weather component (see supplementary material for details).
Our initial guesses for the state parameters and the maximum likelihood estimates of the variance parameters are also given in the supplementary material.

We estimate the $V$, $W_\mu$, $W_\beta$, $W_\psi$, $W_\phi$ and $W_{X_t}$ and the forcing parameters $\varphi$ and $W_\delta$ by maximum likelihood, using the expression for the likelihood provided in \citet{Sansom2018a}.
To ensure the variances estimates were positive, we estimated the log of each parameter.
We recommend using a quasi-Newton method such as the Broyden-Fletcher-Goldfarb-Shanno algorithm for numerical maximization of the likelihood. 
Simplex methods such as Nelder-Mead optimization will tend to be unstable if any of the variances are very close to zero.

\begin{figure*}[t]
  \centerline{\includegraphics[scale=0.5]{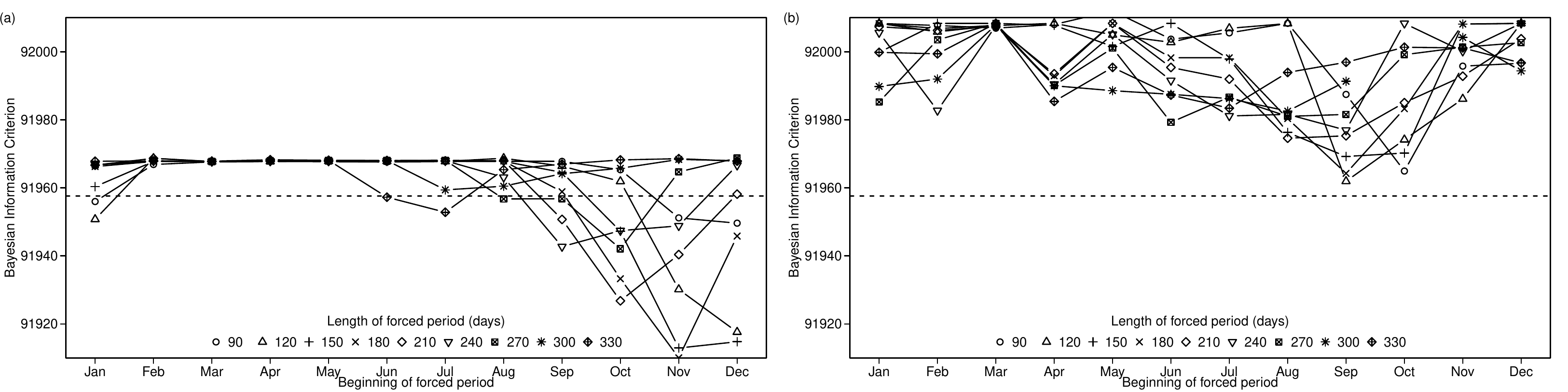}}
  \caption{Choice of influence function $\lambda_t$.
           Bayesian Information Criterion (lower is better) of models with forced periods from 3--11 months starting on the first day of each month were considered.
           The dashed line indicates the model with no predictable signal.
           The influence function was linearly tapered during the first and last 30 days of the forced period, as in Figure~\ref{fig:lambda}.}
  \label{fig:influence}
\end{figure*}

\subsection*{Mean shift or autocorrelation shift?}

Figure~\ref{fig:nao} suggests the presence of external forcing of the NAO between December and March.
In order to discover any other external forcing, a family of 108 different influence functions $\lambda_t$ was constructed.
We considered forced periods starting on the first day of each month, with lengths between 90 and 330 days in 30 day increments, linearly tapered over the first and last 30 days, as in Figure~\ref{fig:lambda}.
Maximum likelihood estimation of the variance and forcing parameters was carried out for each influence function, for both the mean shift model, and the autocorrelation shift model.

Figure~\ref{fig:influence} compares the two families of models using the Bayesian Information Criteria \citep[Chapter 9]{Wilks}.
In the mean shift model, there is no evidence of potentially predictable periods beginning between February and July.
The best models have potentially predictable periods that begin in November or December, and end in March or April.
In comparison, the autocorrelation shift model performs very poorly.
There is little discrimination between the different influence functions, and none outperform a model with no external forcing.
This agrees with the conclusions of \citet{Sansom2018a} who analyzed a shorter NAO index but considered a much larger family of influence functions by Markov Chain Monte Carlo simulation.
We conclude that a temporary mean shift is the most likely explanation for the persistent behavior of the winter NAO, and do not consider the autocorrelation model further.

\begin{figure*}[t]
  \centerline{\includegraphics[scale=0.5]{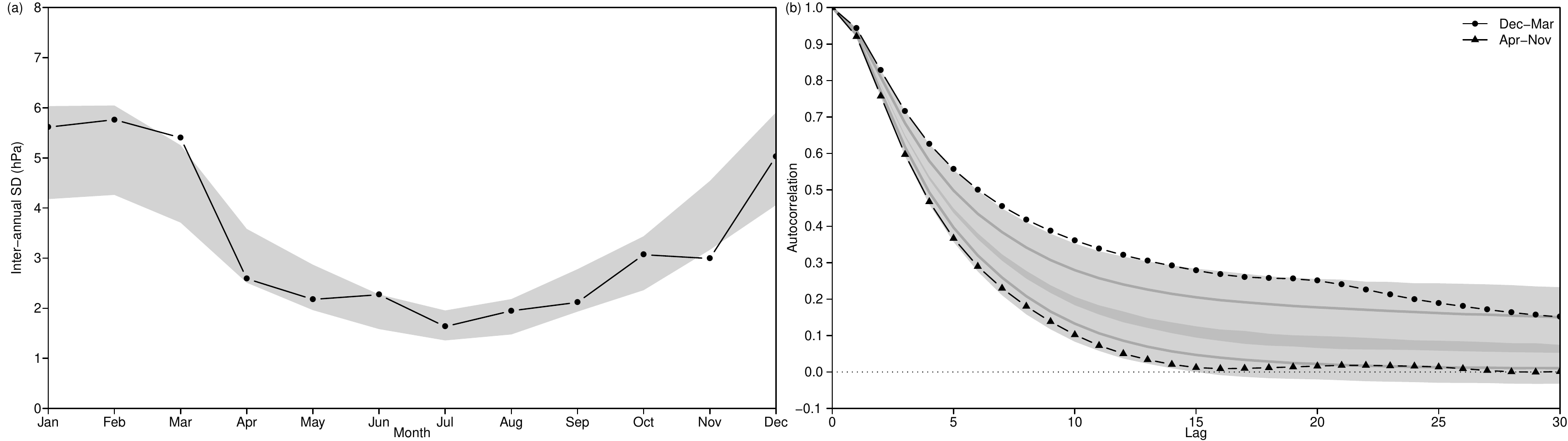}}
  \caption{Posterior predictive checks.
           (a) The inter-annual standard deviation of the monthly mean NAO index, and
           (b) the autocorrelation function of the daily NAO index computed for Dec-Mar and Apr-Nov.
           Black lines represent the sample estimates computed from the observed NAO index as in Figure~\ref{fig:nao}.
           Shaded grey regions represent 95\% credible intervals computed from 1000 simulations of the 50 years between 1968 and 2017 from the statistical model.}
  \label{fig:post}
\end{figure*}

\subsection*{Model checking}

We used simulations of the NAO index between 1968 and 2017 to check whether the mean shift model can adequately reproduce the behavior observed in Figure~\ref{fig:nao}.
The model with $P=5$ struggled to reproduce the summer autocorrelation function (see supplementary material).
We settled on a model with $P=6$ and a 165 day forced period beginning on 1 Nov each year.
The influence function $\lambda_t$ is shown in Figure~\ref{fig:lambda}.

Figure~\ref{fig:post} shows that the chosen model is able to reproduce the observed patterns of both the inter-annual variance and the autocorrelation function.
The pronounced step in the inter-annual variance between December and March is captured reasonably well.
There is also a clear difference between the autocorrelation functions simulated for Dec--Mar and Apr--Nov, although parts of the observed autocorrelation in Dec--Mar lie on the very edge of the simulated 95\% credible interval.
We conclude that a mean shift model is able to explain the observed features of the NAO index.
Simulation studies from the optimal autocorrelation shift model are included in the supplementary material for comparison, however it was unable to reproduce the observed inter-annual variance or autocorrelation function.

\section{Results}
\label{sec:results}

\subsection{Long-term behavior of the NAO index}

\begin{figure*}[t]
  \centerline{\includegraphics[scale=0.5]{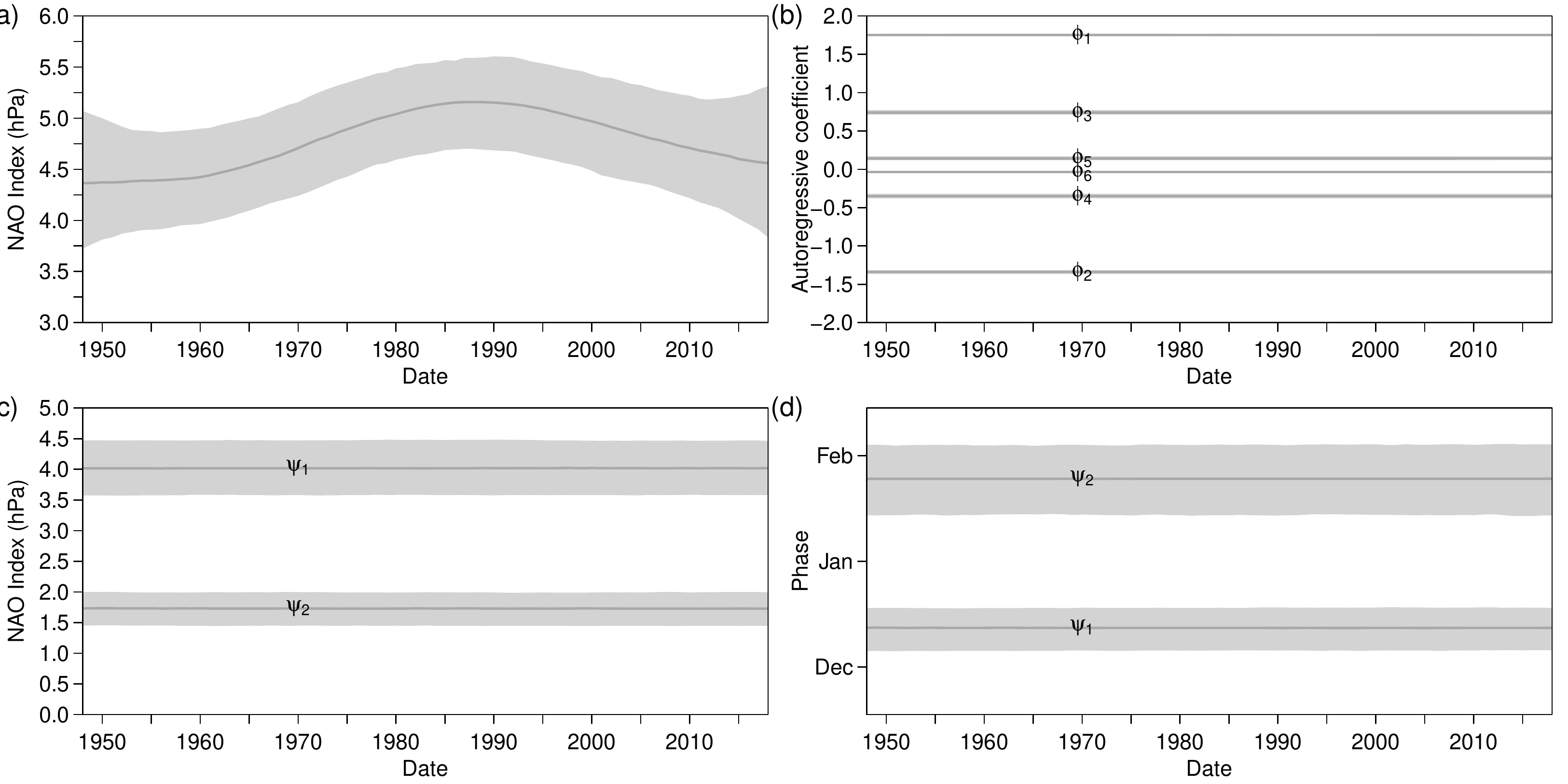}}
  \caption{Long-term trends. 
           (a) The long-term mean component $\mu_t$,
           (b) the autoregressive coefficients $\phi_{1t},\ldots,\phi_{6t}$,
           (c) the amplitude of the annual and semi-annual cycles $\psi_{1t}$ and $\psi_{2t}$,
           (d) the phase of the annual and semi-annual cycles $\psi_{1t}$ and $\psi_{2t}$ plotted as the timing the first peak.
           Solid lines indicate the posterior mean and shading represents \SI{95}{\percent} credible intervals computed from 1000 plausible trajectories for the state parameters $\btheta_1,\ldots,\btheta_T$.
          }  
          \label{fig:trends}
\end{figure*}

The mean component $\mu_t$ is intended to capture gradual changes due to climate change or long-term decadal variability.
Figure~\ref{fig:trends}(a) suggests that the mean of the NAO index has changed gradually over the last 70 years.
The mean $\mu_t$ rose by around \SI{0.8}{\hecto\pascal} (95\% C.I. \SIrange{0.1}{1.3}{\hecto\pascal}) between 1950 and 1990, before declining slightly over the following two decades.
In contrast, the amplitude and phase of the annual and semi-annual cycles in Figures~\ref{fig:trends}(c) and (d) show little sign of variation.
Figure~\ref{fig:trends}(b) also shows very little change in the autoregressive coefficients $\phi_1,\ldots,\phi_6$ over the study period.
This suggests that the seasonal behavior and autocorrelation of the daily NAO index have been effectively constant over the last 70 years, and have not contributed to periods of unusual persistence.

\subsection{Analysis of variance}

\citet{Sansom2018a} describe how standard analysis of variance methods can be applied to our model for the daily NAO index.
Table~\ref{tab:anova} lists the fraction of inter-annual variance explained by each component for the traditional climatological seasons.
The contribution due to observation error $v_t$ is always negligible.
This is due to the fact that our NAO index is computed from reanalysis data.
While there will be errors in the underlying observations, there should be no independent errors between time steps in the reanalysis output.

In DJF, the potentially predictable component explains around \SI{60}{\percent} (\SI{95}{\percent} credible interval \SIrange{48}{70}{\percent}) of the inter-annual variance in the seasonal mean NAO index, compared to \SI{32}{\percent} (\SIrange{24}{40}{\percent}) for the weather component. 
The remaining \SI{8}{\percent} (\SIrange{1}{17}{\percent}) is explained by changes in the mean.
This implies a potential correlation of almost 0.8 for a perfect forecast, similar to the findings of \citet{Scaife2014} and \citet{Athanasiadis2017}.
In JJA there is no external forcing our final model.
Around \SI{87}{\percent} (\SIrange{84}{95}{\percent})of the inter-annual variance is attributed to weather $X_t$.
The remaining \SI{13}{\percent} (\SIrange{5}{16}{\percent})is attributed to slow changes in the mean.

For comparison, we repeated the analysis of \citet{Keeley2009} using the canonical model of \citet{Zwiers1987} on our simple NAO index.
Fixed annual and semi-annual harmonics were estimated by least squares and removed from the data before analysis.
In DJF, the ``best guess'' method of \citet{Keeley2009} suggests that around \SI{75}{\percent} of the inter-annual variance is attributable to external forcing, and \SI{25}{\percent} to weather noise.
These estimates are only just outside of our \SI{95}{\percent} credible intervals.
The difference is partially explained by the contribution due to changes in the mean, which in the method of \citet{Keeley2009} will have been included in the contribution due to external forcing.
In JJA, the \citet{Keeley2009} method estimates that around \SI{84}{\percent} of the inter-annual variance is attributable to weather and \SI{16}{\percent} to external forcing.
It seems that simple potential predictability methods are able to estimate the contribution due to weather noise reasonably accurately, but by definition it cannot distinguish changes in the mean from external forcing.

Using empirical mode decomposition on a related index, \cite{Franzke2011} estimated that around \SI{55}{\percent} of the inter-annual variance in the winter NAO could be explained by external forcing.
This is very similar to our estimate of \SI{60}{\percent}, since empirical mode decomposition also accounts for long-term trends.
\cite{Franzke2011} also estimated that up to \SI{30}{\percent} of the inter-annual variance in summer can also be explained by external forcing.
By fitting a single forcing effect, we have focused on the extended winter period.
Figure~\ref{fig:influence} suggests that there may also be short period of limited predictability in summer, however further investigation is beyond the scope of this study.

\begin{table*}[t]
  \caption{Analysis of variance. Bracketed values indicate 95\% credible intervals.}
  \label{tab:anova}
  \centering
  \begin{small}
    \fbox{\begin{tabular}{lcccc}
      \centering 
       & Mean & External & Weather & Error \\
       MAM & 0.05 (0.00,0.11) & 0.35 (0.22,0.46) 
           & 0.60 (0.49,0.71) & 0.00 (0.00,0.00) \\
       JJA & 0.13 (0.05,0.16) & 0.00 (0.00,0.00)
           & 0.87 (0.84,0.95) & 0.00 (0.00,0.00) \\
       SON & 0.03 (0.01,0.05) & 0.11 (0.04,0.20) 
           & 0.86 (0.77,0.93) & 0.00 (0.00,0.00) \\
       DJF & 0.08 (0.01,0.17) & 0.60 (0.48,0.70) 
           & 0.32 (0.24,0.40) & 0.00 (0.00,0.00) \\
    \end{tabular}}
  \end{small}
\end{table*}

\subsection{What happened in particular years?}

\begin{figure*}[t]
  \centerline{\includegraphics{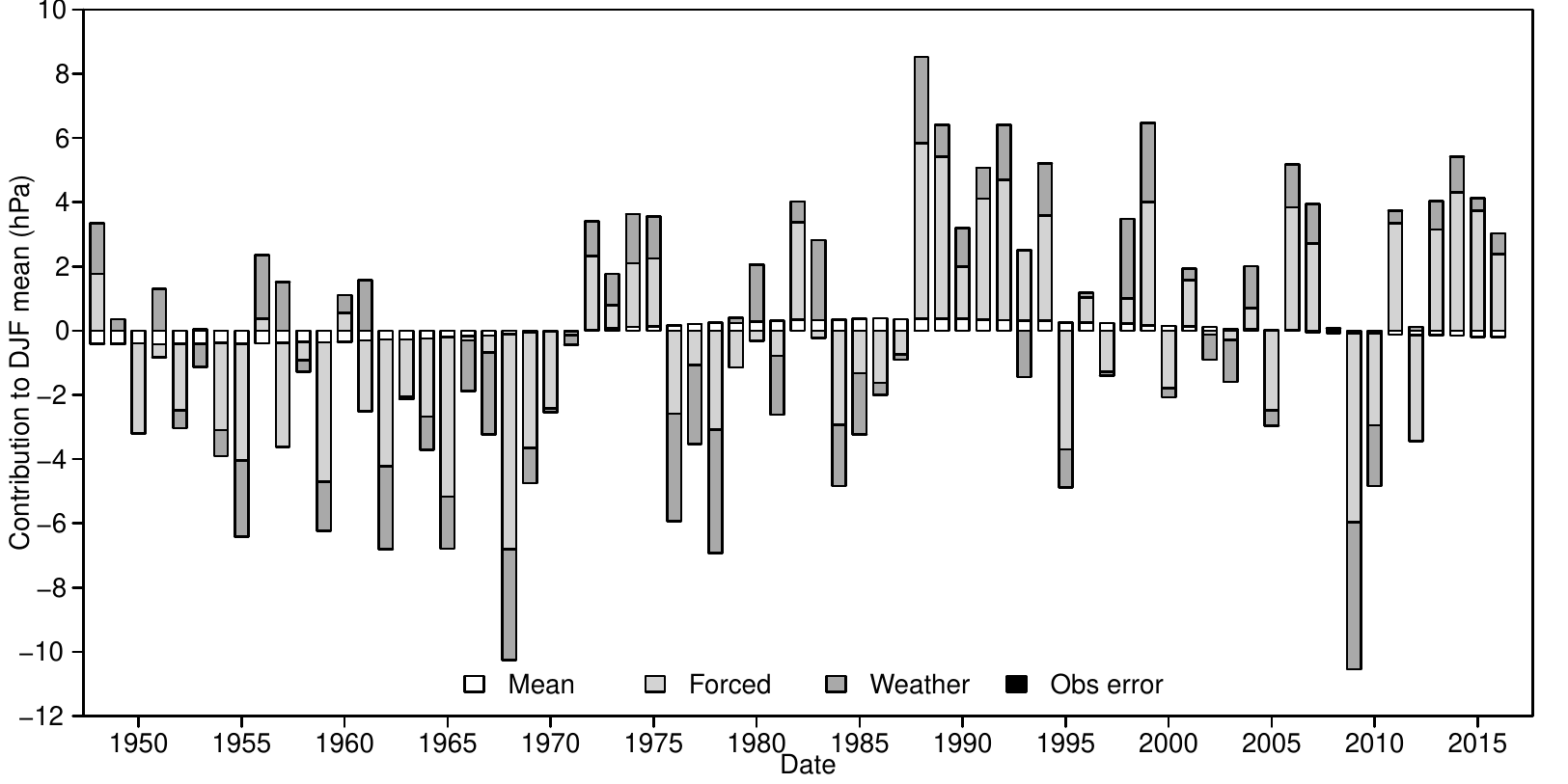}}
  \caption{The posterior mean contribution of the mean, potentially predictable, weather and observation error components to the DJF mean NAO index every year between 1958 and 2017, computed from 1000 plausible trajectories for the state parameters.}
  \label{fig:attribution}
\end{figure*}

Figure~\ref{fig:attribution} shows the estimated contributions of the mean $\mu_t$, potentially predictable $Z_t$, weather $X_t$, and observation error $v_t$ components of the model to the winter (DJF) mean NAO index during each year of the study period.
The slow increase to slight decrease in the long-term mean $\mu_t$ of the NAO is clearly visible in the white bars corresponding the mean component $\mu_t$.
The contributions of the forced $Z_t$ and weather $X_t$ components appear positively correlated ($\rho=0.69$), despite being independent in the model.
The estimates presented in Fig.~\ref{fig:attribution} are a summary of 1000 plausible trajectories $t=1,\ldots,T$ for each component.
The positive correlation is an anomaly caused by summarizing over the sample of plausible trajectories.
If the forcing effect $\delta_t$ is too weak to distinguish in a particular year, then on average the seasonal mean will be partitioned roughly in line with the analysis of variance in Tab.~\ref{tab:anova}, leading to correlated estimates of the the forced $Z_t$ and weather $X_t$ components.
The fact that the relative contributions of the forced and weather components varies significantly, and in some years have different signs (e.g., 1961), indicates that the model is able to distinguish the two effects.
If the two components were actually correlated, then we would expect the seasonal means of each component from each individual trajectory to also be correlated.
This is not the case, performing the same attribution analysis for each individual trajectory returns an average correlation of $\rho = 0.02$ $(-0.13,+0.21)$.

\begin{figure*}[t]
  \centerline{\includegraphics{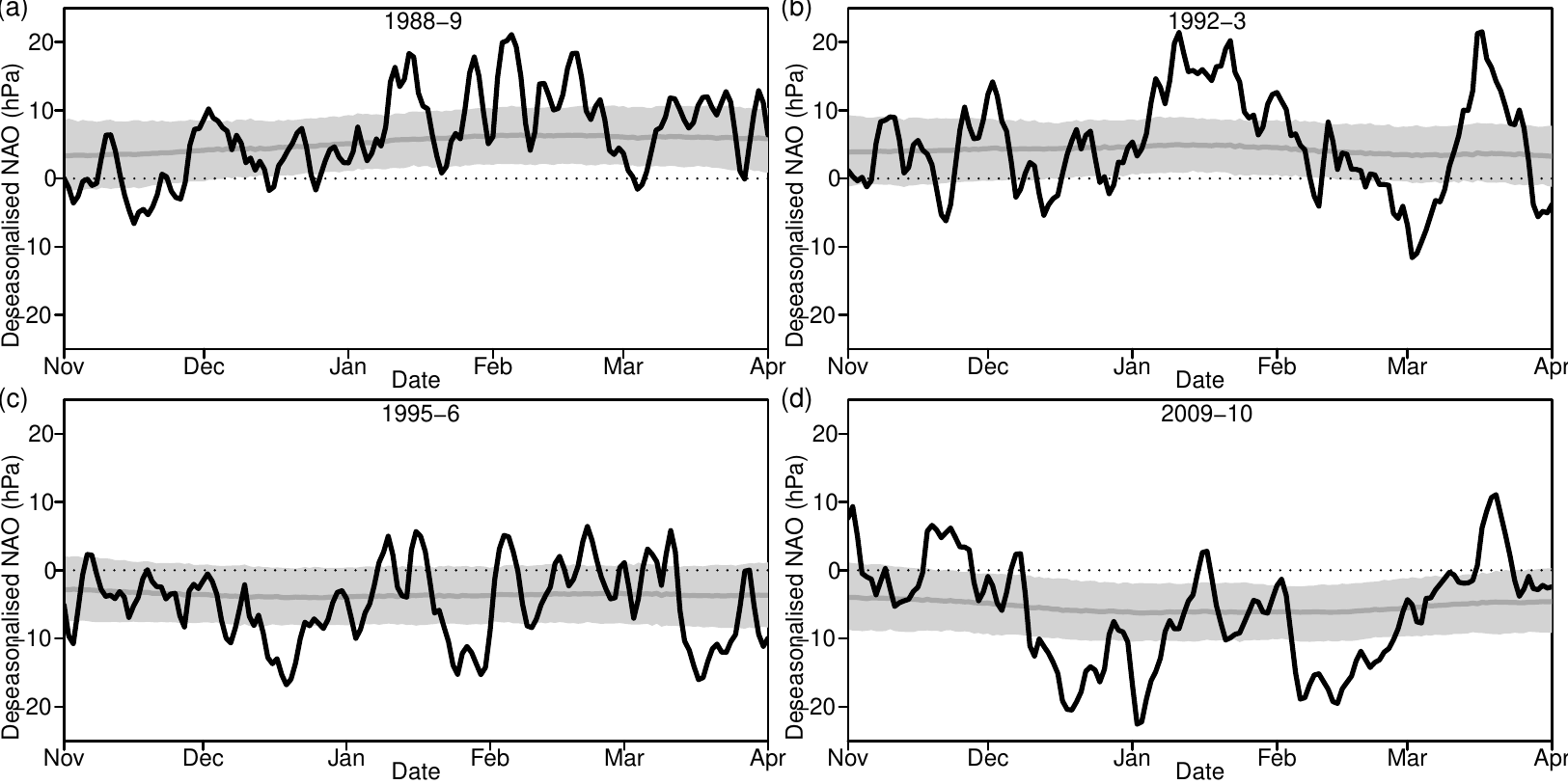}}
  \caption{The evolution of the forcing effect $\delta_t$ during the winters of 1988--89, 1992--93, 1995--96 and 2009--10.
           Solid gray lines indicate the mean of 1000 plausible trajectories for the forcing effect $\delta_t$.
           Shaded regions indicate 95\% credible intervals for $\delta_t$ computed from the same samples.
           Solid black lines are the deseasonalized observations for comparison.}
  \label{fig:years}
\end{figure*}

Figure~\ref{fig:years} shows the estimated evolution of the forcing effect $\delta_t$ during four recent winters with unusually strong winter NAO anomalies.
Previous studies assumed that any external forcing was constant throughout a particular season \citep[e.g.,][]{Keeley2009}.
Our analysis suggests that the assumption of constant forcing is not justified.
In 1989--90, the forcing effect increased in strength throughout December and January,  reaching a peak during February.
February 1990 was marked by a series of eight strong extra-tropical cyclones impacting Europe.
The strong positive NAO in 1989--90 has been linked to a strong La Ni\~{n}a event \citep[e.g.,][]{Bronnimann2006}.
In 1992--93 and 1995--96 the forcing effect remained fairly constant throughout the winter season.
In 2009--10 the forcing effect started out only moderately negative during November before increasing in strength during December, January and February.
The strong negative NAO in 2009-10 has been linked to a strong El Ni\~{n}o event and an associated sudden stratospheric warming \citep{Fereday2012,Scaife2016a}.

\subsection{Seasonal predictability}
\label{sec:forecasting}

\begin{figure*}[t]
  \centerline{\includegraphics[scale=0.5]{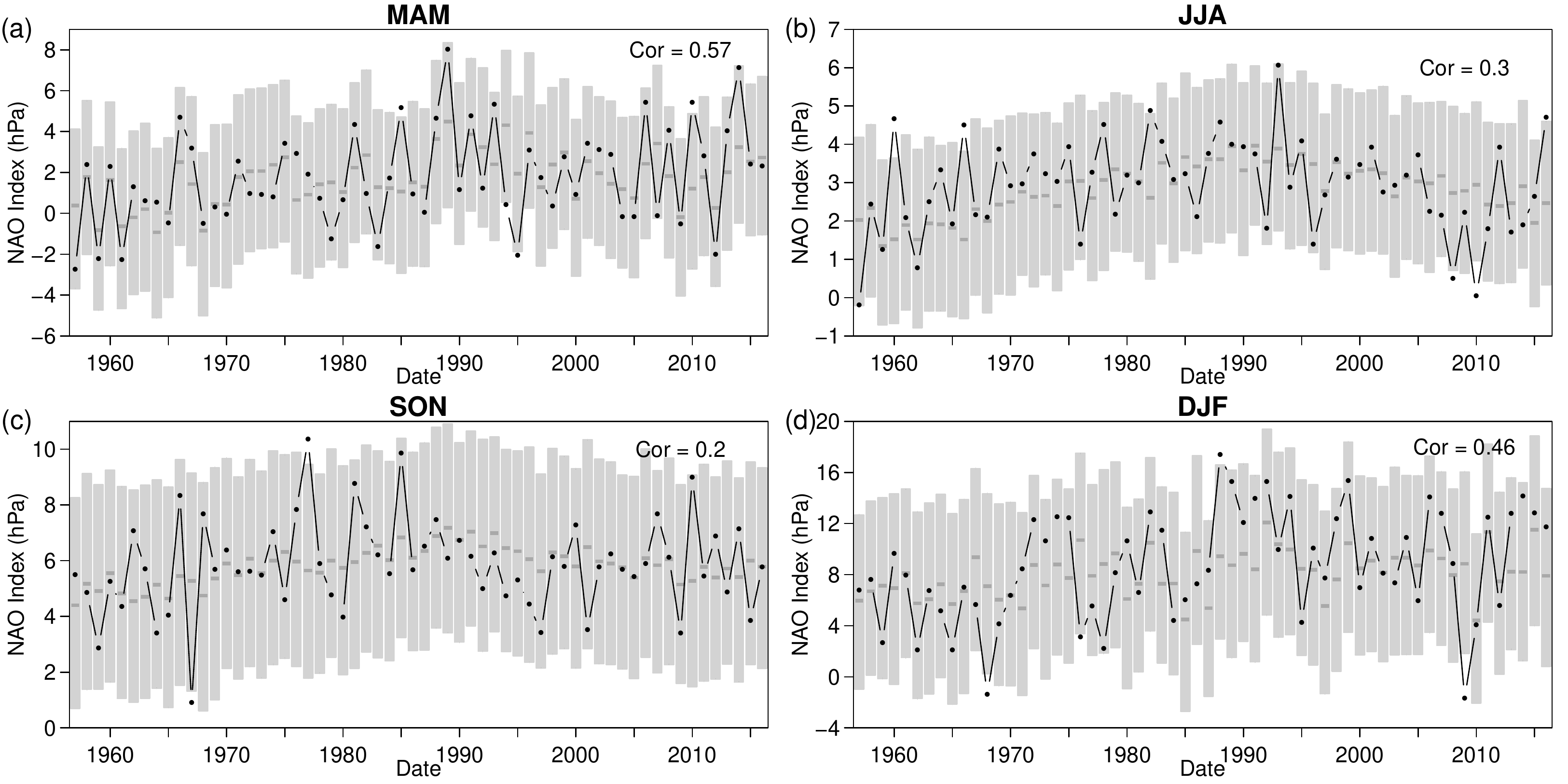}}
  \caption{Seasonal forecasting.
           1000 forecast members were initialized on 1 March (MAM), 1 June (JJA), 1 September (SON) and 1 December (DJF) every year between 1957 and 2016 and run forwards for three months.
           Dark gray points indicate the seasonal forecast means.
           Light gray shaded regions represent the 95\% prediction intervals.
           The observed seasoanl mean of the NAO index is plotted in black for comparison.}
  \label{fig:forecasts}
\end{figure*}

Between December and March, the forcing effect $\delta_t$ represents a slowly varying shift in the mean of the NAO index.
If we can estimate the external forcing early in the season, then we should have useful predictability for the rest of the winter.
Figure~\ref{fig:forecasts} shows the results of forecast tests for the seasonal mean NAO index.
1000 forecast members were initialized on the first day of each season (1 March, 1 June, 1 September and 1 December) for every year between 1957 and 2016 and propagated forward until the end of that season.
This differs from standard practice in seasonal forecasting where forecasts are usually initialized 30 days before the beginning of the season.
However, since the forced period selected in Section~\ref{sec:modeling} only begins on 1 November, the model requires time to detect the forced signal.
For the DJF forecasts, the model only has the 30 day transition period from unforced to forced during November to learn the forcing effect.
The forecast mean achieves a correlation of 0.48 with the observations of the Dec-Jan-Feb winter mean, despite the limited time to learn the forcing effect.
The statistical model notably fails to predict the extreme winter of 2009-10.
However, in Figure~\ref{fig:years}(d) we see that the NAO index did not become unusually negative until mid-December, there was no signal for the model to identify in November.

It would be a mistake to regard the seasonal forecasts from the statistical model as simple persistence forecasts.
Extraction of the predictable signal depends on the day-to-day variation in the observed index, not simply the mean of the index in recent observations.
For comparison, optimal linearly and exponentially weighted persistence forecasts were computed for the DJF mean NAO index over the same period (see supplementary material for details).
The correlations between observations and the exponentially and linearly weighted forecasts are 0.35 and 0.37 respectively.
The statistical model outperforms both persistence forecasts.

The forcing effect $\delta_t$ does not act on the NAO at all during summer (JJA), and only weakly during the final month of autumn (SON).
The weak positive correlation between the forecasts and observations in JJA and SON is due to the long-term trend in Figure~\ref{fig:trends}(a), which is clearly visible in both the forecasts and observations in Figure~\ref{fig:forecasts}.
Since the external forcing appears to influence the NAO index during March and April, we might also expect to have useful predictability for the spring season (MAM).
The MAM forecasts in Figure~\ref{fig:forecasts} achieve a correlation with the observations of 0.57, making them more skillful than the statistical forecasts for Dec-Jan-Feb.
This may appear to contradict the analysis of variance in Table~\ref{tab:anova} which indicated that a greater proportion of the inter-annual variance was explained by external forcing in winter than in spring.
However, by the end of February the model has assimilated three months of additional data with which to estimate the forcing effect.
Therefore, the estimate of the forced component is more precise and the forecast becomes more accurate.
\citet{Jia2017} and \citet{Saito2017} also found increased skill for MAM forecasts compared to DJF or JFM forecasts.

\begin{figure}[t]
  \centerline{\includegraphics[scale=0.5]{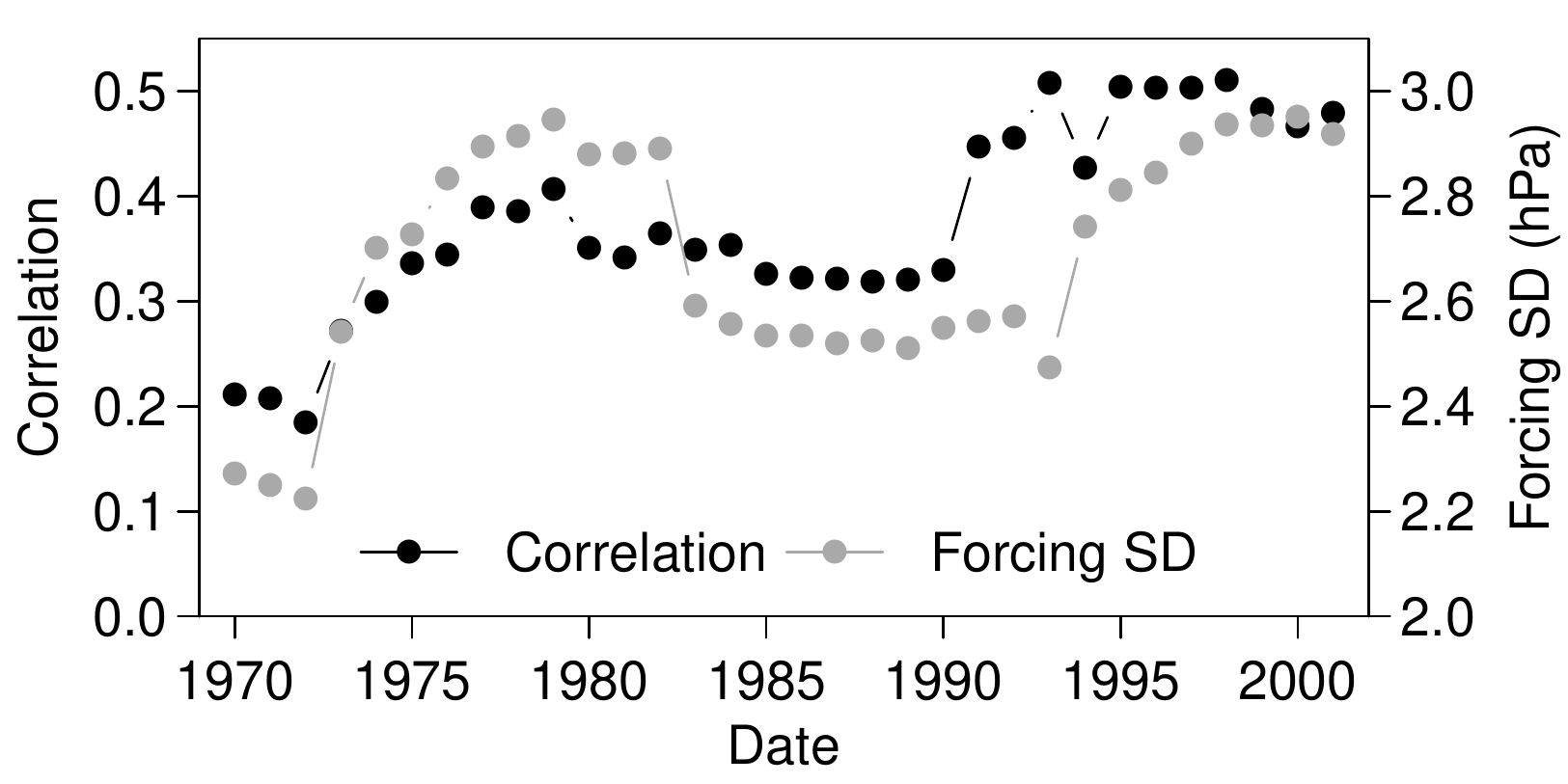}}
  \caption{Variation in forecast skill. 
           Correlation between forecasts and observations of DJF mean NAO index, computed for 30-year moving windows between 1956 and 2016.
           Values are plotted at the 15th year of each window.
           The inter-annual standard deviation of the DJF mean estimates of the forcing effect $\delta$ from Figure~\ref{fig:attribution} are plotted for the same 30-year windows for comparison.
           }
  \label{fig:correlation}
\end{figure}

\citet{Weisheimer2017} noted increased correlation between forecasts of the NAO and observations between 1970--2000 compared to 1950--70.
Figure~\ref{fig:correlation} shows correlations for 30-year windows centered on 1970--2001 based on the statistical forecasts in Figure~\ref{fig:forecasts}.
The statistical forecasts show a similar pattern of increasing correlation to that of \citet[Figure 2a]{Weisheimer2017}.
The pattern of changing skill correlates well with changes in the inter-annual standard deviation of the seasonal mean of the forcing effect $\delta$, also shown in Figure~\ref{fig:correlation}.
This relationship makes physical sense since if the apparent variability of the forcing effect increases, then so will the signal-to-noise ratio of the predictable signal.
However, as in \citet{Weisheimer2017}, we note that the difference in skill between the early 1970s and mid 1990s is not significant.

\begin{figure}[t]
  \centerline{\includegraphics[scale=0.5]{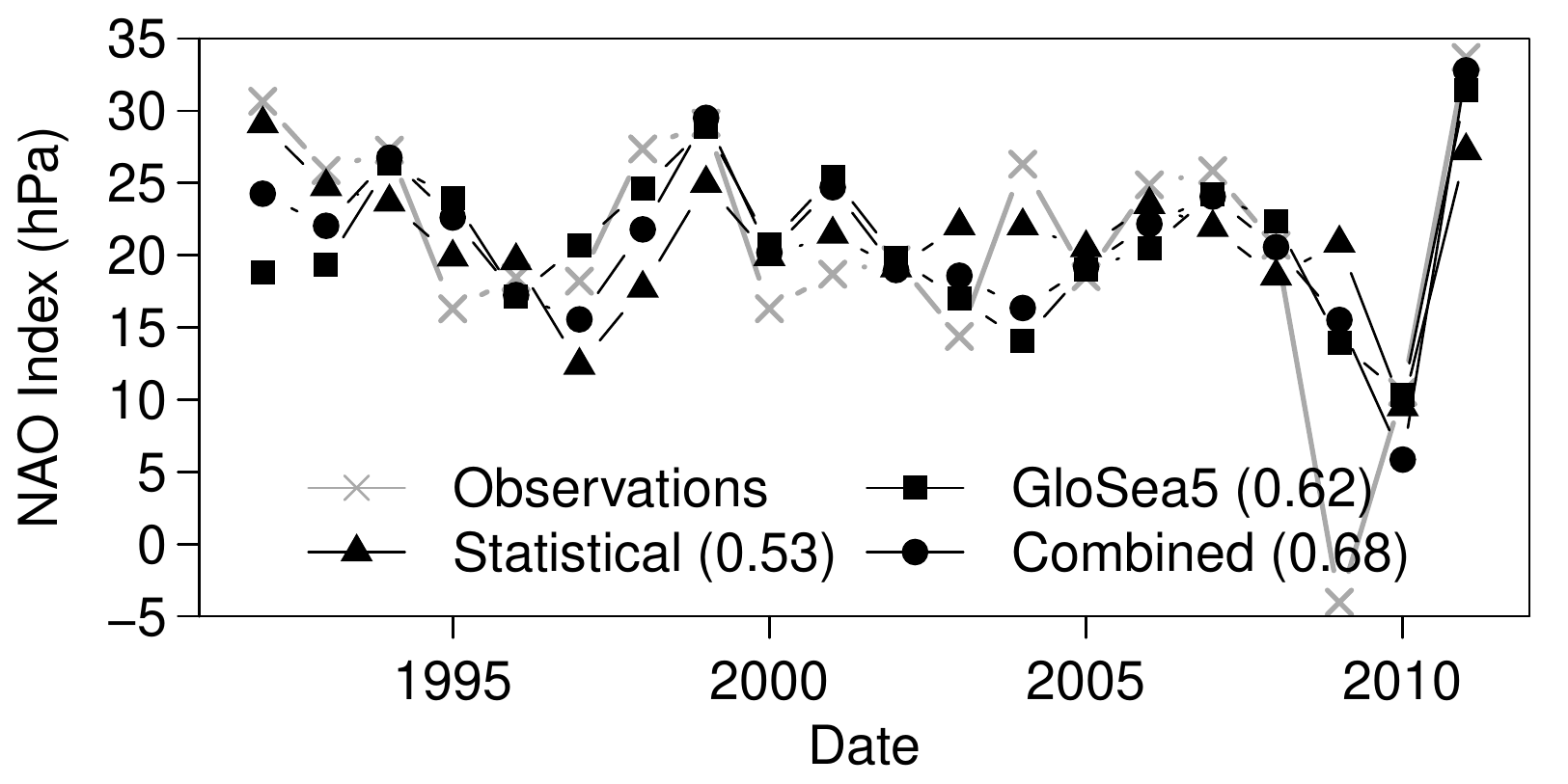}}
  \caption{Comparison with GloSea5.
           Ensemble mean forecasts for DJF each year between 1992 and 2011 from the statistical model and GloSea5.
           Both sets of forecasts have been recalibrated by linear regression to make them comparable.
           The GloSea5 verifying observations are plotted for comparison.
           A combined forecast produced by multiple linear regression of the two sets of ensemble means on the observations is also plotted.}
  \label{fig:glosea}
\end{figure}

Figure~\ref{fig:glosea} compares the seasonal forecasts for the winter season (Dec-Jan-Feb) with the state-of-the art GloSea5 seasonal prediction system for the years 1992--2011.
To facilitate the comparison, the ensemble means of both sets of forecasts were recalibrated using linear regression to be compatible with the verifying observations.
The statistical forecasts achieve a correlation of 0.53 with the verifying observations, compared to 0.62 for GloSea5.
The GloSea5 forecasts are initialized one month earlier, but have the advantage of assimilating data from the entire climate system.
In contrast, the statistical model must learn the forcing effect based on only 30 days of NAO index observations.
Figure~\ref{fig:glosea} also illustrates a combined forecast using both the GloSea5 and statistical forecasts computed by multiple linear regression of the ensemble means on the observations.
The correlation of the combined forecasts with the observations increases to 0.68, suggesting the potential to cheaply update and improve dynamical forecasts with new data using statistical forecasts.

\section{Discussion}
\label{sec:discussion}

In this study, we applied state-space modeling techniques to assess persistence and predictability in climate indices.
Externally forced predictable signals can be separated from slow changes in the mean and rapid changes due to day-to-day weather.
Time series can be analyzed without splitting into seasons since non-stationary mean and seasonal components are estimated at the same time as the predictable signal.
We also consider changes in the temporal dependence structure of the index.
Our methodology can distinguish predictability due to temporary changes in the mean of an index, from temporary changes in the day-to-day persistence of an index.
If the timing of the predictable signal is not known, it can be extracted from the data using simple model comparison methods.
The individual components of the model have clear physical interpretations, e.g., annual mean, long-term trend, annual cycle, semi-annual cycle etc.
This makes it easy to incorporate physical knowledge and intuition through choices in the model structure (e.g., number of harmonic components), initial guesses for component values, and the choices for the model variances.

Existing methods for assessing potential predictability in climate indices only estimate the fractions of inter-annual variance explained by a potentially predictable signal and accumulated weather noise.
The methodology demonstrated here provides a wealth of additional information.
The variance explained by observation errors and changes in the mean and annual cycle can also be quantified.
The state of each component is estimated at every time point, including corresponding uncertainty estimates.
Therefore, we can estimate how each component has evolved over the study period.
The contribution of each component to individual seasonal means can be also estimated, in addition to the overall variance explained.
The predictable signal is treated as a dynamic process, rather than fixed throughout a season, allowing us to estimate changes in the forced signal during a particular season.
The same model can also be used for forecasting and can be tested without the need for cross validation.

Our analysis of the daily NAO index suggests a distinct split in the dynamics of the NAO between December, January, February, March and the rest of the year.
Focusing on the traditional DJF period, around \SI{60}{\percent} of the inter-annual variance in the winter mean NAO is attributable to external forcing, \SI{32}{\percent} to weather noise and \SI{8}{\percent} to long-term trends.
Our estimate of the contribution of external forcing is similar to that of \citet{Franzke2011}, but lower than that of \citet{Keeley2009}, in part due to the separation of long-term trends from the forced signal.
Skillful statistical forecasts of the DJF mean NAO are possible from the end of November and achieve a correlation with the observations of 0.48.
Our analysis indicates that external forcing, and hence predictability, extend into March.
We found little evidence of potentially predictable signals outside of the extended winter season, although a weak signal may exist for part of boreal summer.

It is striking that the statistical forecast model is able to reproduce the apparent time-varying forecast skill in the NAO noted by \cite{Weisheimer2017}.
Dynamical models might be affected by more limited observations of the ocean state in the middle compared to the end of the century, leading to less precise initialization and less skillful forecasts.
The statistical model only assimilates observations of the NAO itself, which is likely to be less strongly affected by changes in the observation network, due to the large area the index is averaged over.
This suggests that the time-varying predictability may be a physical property of the system.
It is also interesting that the statistical model is able to approach the skill of state-of-the-art seasonal forecasting model, although with the caveat of a shorter lead time.
Our results suggest possibilities for updating expensive dynamical seasonal forecasts using inexpensive statistical forecasts.
Such a hybrid approach might be used to further increase skill through dynamic ensemble design, with additional ensemble members being commissioned dependent on the conditions indicated by the statistical forecasts.

There is a continuing debate over the signal-to-noise ratio of the NAO in dynamical forecast models and its effect on achievable forecast skill \citep{Eade2014,Shi2015}.
One of the strengths of the methodology described here is its ability to break down time series of climate indices into easily interpretable components.
Fitting the statistical model proposed here to long running simulations from dynamical models would enable detailed comparisons between the dynamics of the models and the Earth system.
Such detailed benchmarking might lead to new insights into the performance of dynamical models suggest areas for further development.
With additional development it may be possible to apply a similar analysis to forecast or hindcast datasets for a more direct comparison.

\section*{Acknowledgments}
The authors gratefully acknowledge the support of the Natural Environment Research Council grant NE/M006123/1, and Adam Scaife of the UK Met Office for providing the GloSea5 hindcast dataset and for helpful comments on an earlier version of this manuscript.

\bibliographystyle{abbrvnat}
\bibliography{library}

\end{document}


\maketitle

\section*{\centering Supplementary material}

\section{Exploratory analysis}
\label{sec:exploratory}

\begin{figure}[t]
  \centerline{\includegraphics{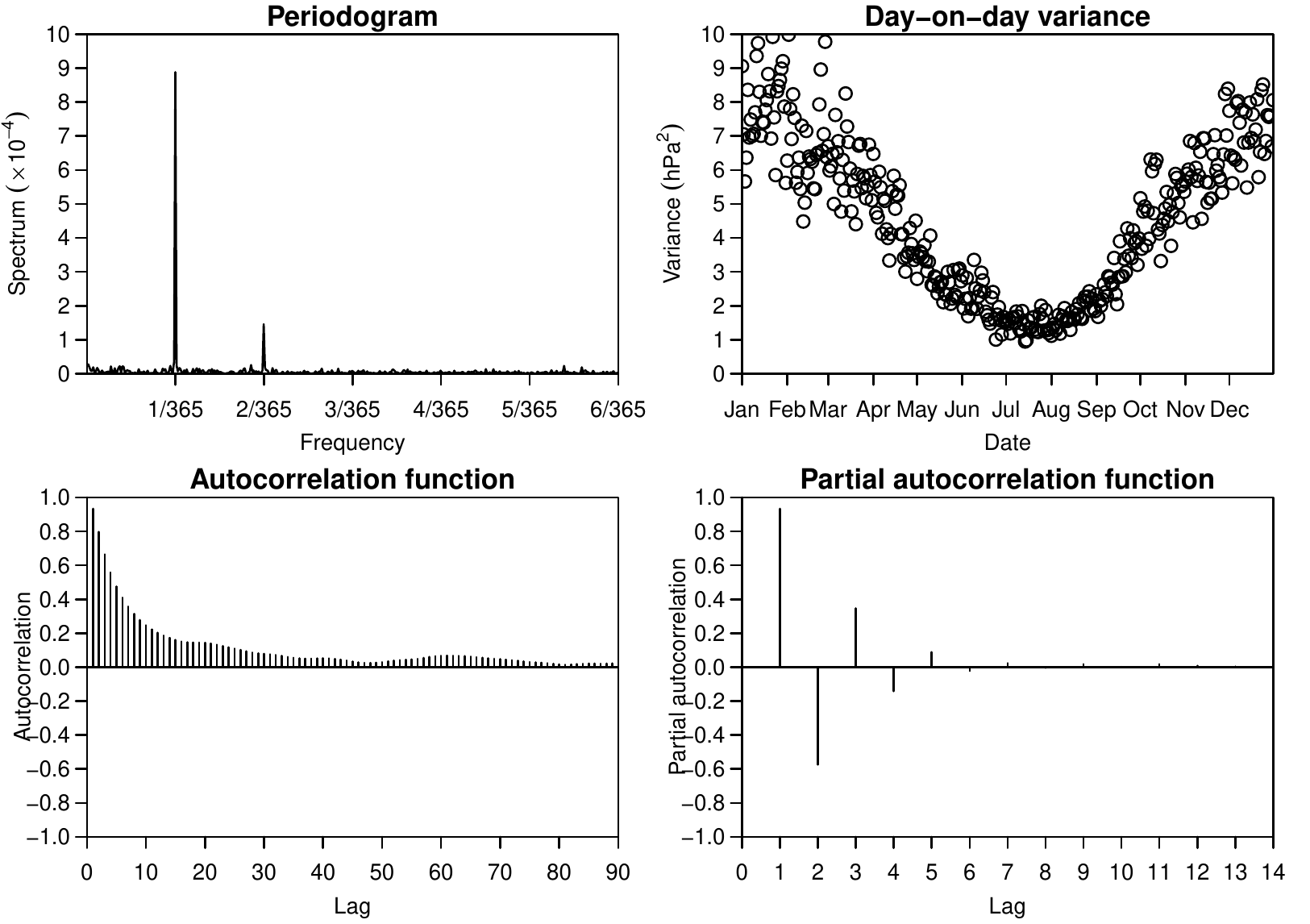}}
  \caption{Exploratory analysis.
           (a) Periodogram,
           (b) day-on-day variance of first differences,
           (c) autocorrelation function, and
           (d) partial autocorrelation function
           of the daily NAO index.}
  \label{fig:exploratory}
\end{figure}

The aim of the exploratory analysis is to identify the number of harmonic components $K$ required to represent any annual cycle in the observations, the order $P$ of the autoregressive process required to represent the weather, and any other features that may need to be included in the model.
The periodogram in Figure~\ref{fig:exploratory}(a) suggests that the NAO index contains annual and semi-annual cycles, i.e., $K = 2$.

The observed index was linearly detrended and fixed annual and semi-annual cycles removed by linear regression before examining the autocorrelation structure.
The autocorrelation and partial autocorrelation functions in Figures~\ref{fig:exploratory}(c) and (d) both decay rapidly, indicating that an autoregressive process is a good representation of the day-to-day variability .
The partial autocorrelation function is approximately zero at lags greater than five, suggesting an AR(5) process for the weather variability, i.e., $P = 5$.

Figure~\ref{fig:exploratory}(b) shows the inter-annual variance in the first differences (e.g., $y_t - y_{t-1}$) of the NAO index computed for each day of the year.
There is a clear annual cycle in the day-to-day variability of the NAO index.
The day-to-day variability is represented in our model by the variance $W_X$.
We model the annual cycle in the day-to-day variance as
\begin{align*}
  W_{X t} & = W_X + \sqrt{a^2 + b^2}+ a \sin \omega t + b \cos \omega t &
    W_X > 0.
\end{align*}
The parameters $W_X$, $a$ and $b$ are assumed to be constant and are estimated by maximum likelihood with the other variance parameters.

\section{Initial guesses for state variables}
\label{sec:priors}

Our initial guesses for the state variables $\mu_t$, $\beta_t$, $\psi_{1 t},\psi_{1 t}^\star,\ldots,\psi_{K t},\psi_{K t}^\star$, $X_{t},\ldots,X_{t-P+1}$,$\phi_{1 t},\ldots,\phi_{P t}$ at time $t=0$ are shown in Table~\ref{tab:priors}.
The initial guess for $mu_t$ at $t=0$ is based on Figure~7 of \citet{Hsu1976} which suggests an overall mean of \SIrange{4}{8}{\hecto\pascal}.
The prior on the local trend $\beta_t$ is based on our judgment that the NAO mean is unlikely experience a local change equivalent to more than 1 hPa/yr.
The analysis of \citet{Chen2012} suggests an annual cycle of approximately \SI{4}{\hecto\pascal} and a semi-annual cycle of around \SI{2}{\hecto\pascal}.
The daily NAO index in Figure~1(a) of the main study has a range of approximately \SI{40}{\hecto\pascal}.
Therefore, the TVAR residuals $X_0,\ldots,X_{1-p}$ were assigned independent normal priors mean \SI{0}{\hecto\pascal} and variance $10^2$.
\citet{Masala2015} fitted an AR(5) model to NAO data and found coefficients of alternating sign in the range (-1.3,+1.8).
Therefore, the TVAR coefficients $\phi_{1 0},\ldots,\phi_{p 0}$ were assigned independent normal priors with means given by the estimates of \citet{Masala2015} and variance $0.2^2$.

The inter-annual standard deviation of the winter mean NAO index is approximately \SI{5}{\hecto\pascal} according to Figure~1 of the main study.
Therefore, our initial guess for the forcing effect $\delta_t$ at $t=0$ in the mean effect model is \SI{0}{\hecto\pascal}, with variance $5^2$.
The change in the autoregressive coefficients under the autocorrelation effect model is not expected to be large.
Therefore, our initial guess for the forcing effects $\delta_{p t}$ at time $t=0$ in the autocorrelation model has mean 0 and variance $0.2^2$

\begin{table}[t]
  \begin{center}
    \begin{small}
      \begin{tabular}{lccc}
        \hline \hline
        \centering 
        Component & Parameter & Mean & Variance \\
        \hline
        Mean                  & $\mu$                
                              & 6 hPa     & $2^2$       \\
        Local trend           & $\beta$              
                              & 0 hPa/yr  & $0.002^2$    \\
        Annual cycle          & $\psi_1,\psi_1^\star$ 
                              & 0 hPa     & $3^2$     \\
        Semi-annual cycle     & $\psi_2,\psi_2^\star$               
                              & 0 hPa     & $2^2$     \\
        Weather noise         & $X_t,\ldots,X_{t-p+1}$ 
                              & 0 hPa     & $10^2$      \\
        AR coefficients       & $\phi_1,\ldots,\phi_p$
                              & 0         & $1^2$       \\        
        \hline
      \end{tabular}
    \end{small}
  \end{center}
  \caption{Initial guesses for the state variables.
           The initial guesses are normally distributed and express our prior uncertainty about the value of the state variables at time $t=0$.}
  \label{tab:priors}
\end{table}

\section{Estimation of the variance parameters}
\label{sec:ml}

The variance parameters $V$, $W_\mu$, $W_\beta$, $W_\psi$, $\sigma_X^2$, $a_X$, $b_X$, $W_\phi$, $W_\delta$ and the coefficient $\varphi$ must be estimated in order to fit the model.
The model fitting procedures developed by \citet{Sansom2018a} include efficient calculation of the likelihood.
Therefore, the unknown parameters can be easily estimated by maximum likelihood.
Numerical optimization of the likelihood was performed using the Broyden-Fletcher-Goldfarb-Shanno algorithm as implemented in the R programming language.
For parsimony, the variances $W_\mu$ and $W_\psi$ were assumed to be equal since.
In order to enforce the constraint that the variance parameters are all positive, optimization was performed on the log scale for $V$, $W_\mu$, $W_\beta$, $W_X$, $W_\phi$, $W_\delta$.
The coefficient $\varphi$ is expected to lie in the range $(0,1)$ so was fitted on a logistic scale.
The maximum likelihood estimates of the parameters for the final model with $K=2$ and $P=6$ are shown in Table~\ref{tab:variances}

\begin{table}[t]
  \begin{center}
    \begin{small}
      \begin{tabular}{lcc}
        \hline \hline
        \centering 
        Component & Parameter & Estimate \\
        \hline
        Long-term mean       & $W_\mu$    & $3.5 \times 10^{-8}$  \\
        Local trend          & $W_\beta$  & $2.8 \times 10^{-12}$ \\
        Forcing coefficient  & $\varphi$  & 0.995                 \\
        Forcing variance     & $W_\delta$ & $1.3 \times 10^{-1}$  \\
        AR coefficients      & $W_\phi$   & $6.1 \times 10^{-12}$ \\
        Weather noise        & $W_X$      &  2.39                 \\
                             & $a_X$      &  0.39                 \\
                             & $b_X$      &  1.64                 \\
        Observation error    & $V$        & $2.5 \times 10^{-5}$ \\
        \hline
      \end{tabular}
    \end{small}
  \end{center}
  \caption{Maximum likelihood estimates of the model parameters.}
  \label{tab:variances}
\end{table}

\section{Posterior predictive checks}
\label{sec:postpred}

\begin{figure}[t]
  \centerline{\includegraphics[scale=0.5]{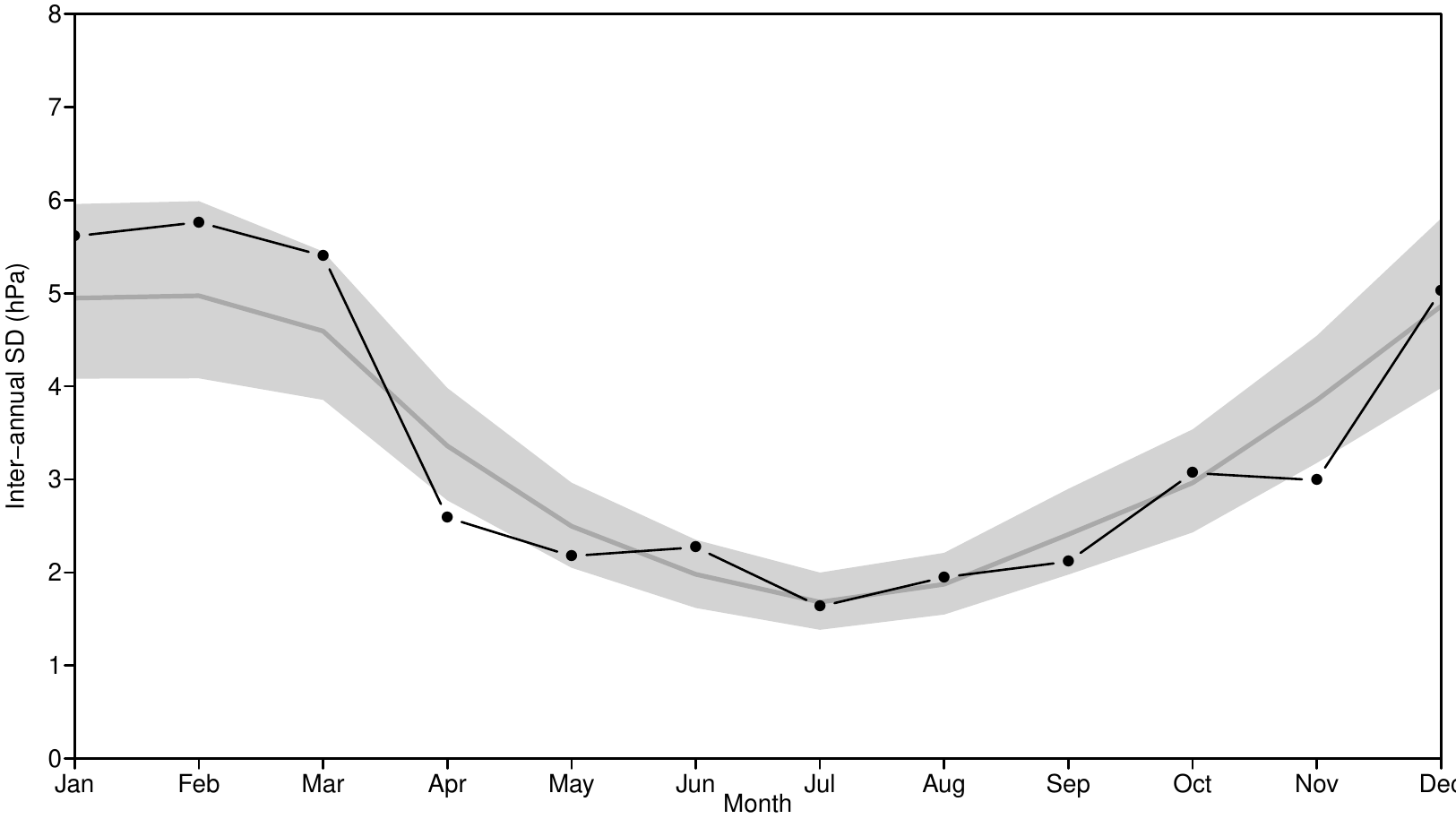}
              \includegraphics[scale=0.5]{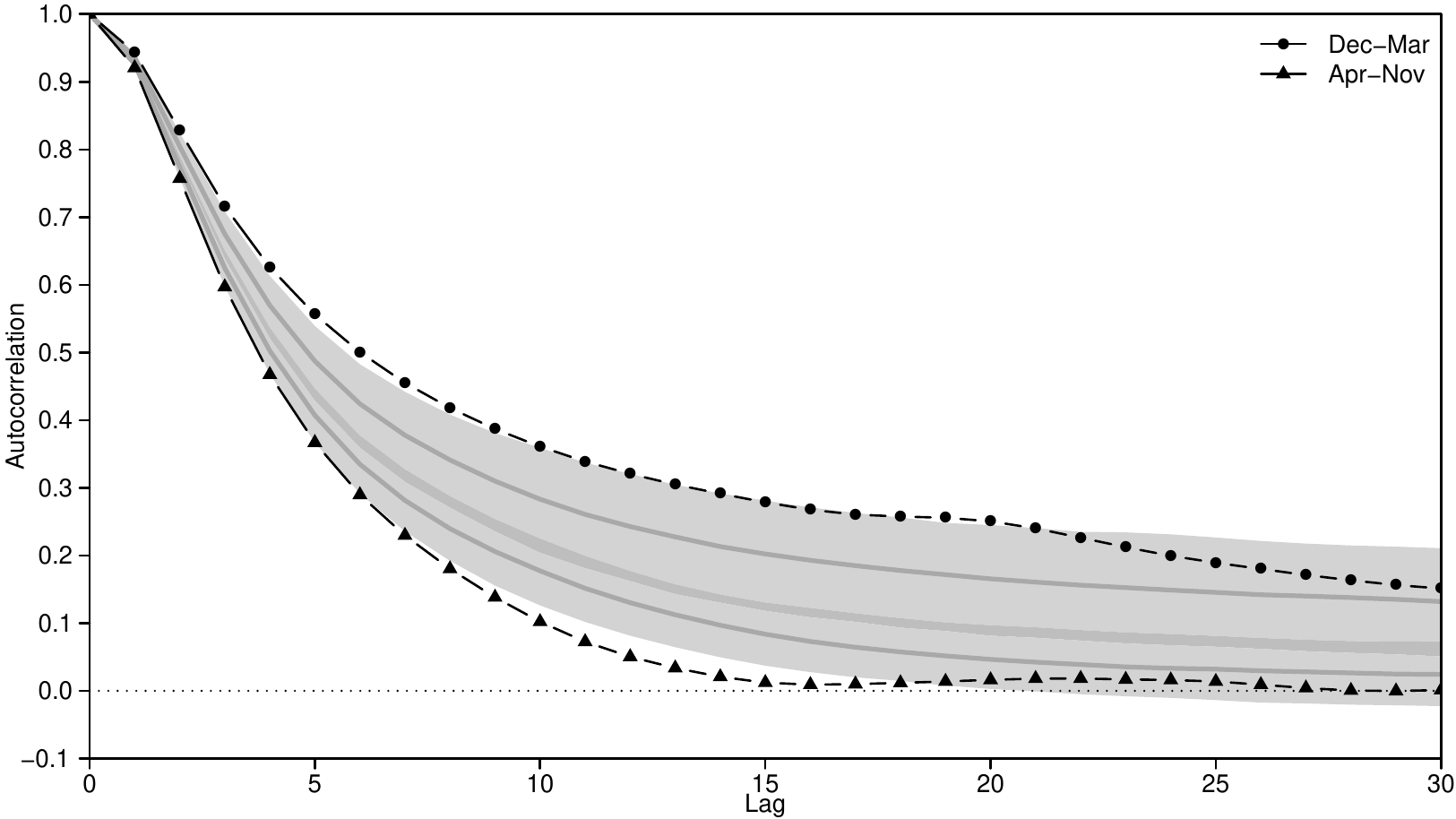}}
  \caption{Posterior predictive checks.
           (a) The inter-annual standard deviation of the monthly mean NAO index, and 
           (b) the autocorrelation
function of the daily NAO index computed for Dec-Mar and Apr-Nov. 
Black lines represent the sample estimates computed from the observed
NAO index as in Figure~1 of the main text. 
Shaded grey regions represent 95\% credible intervals computed from 1000 simulations of the 50 years between 1968
and 2017 from the statistical model.}
  \label{fig:postpred}
\end{figure}

\begin{figure}[t]
  \centerline{\includegraphics[scale=0.5]{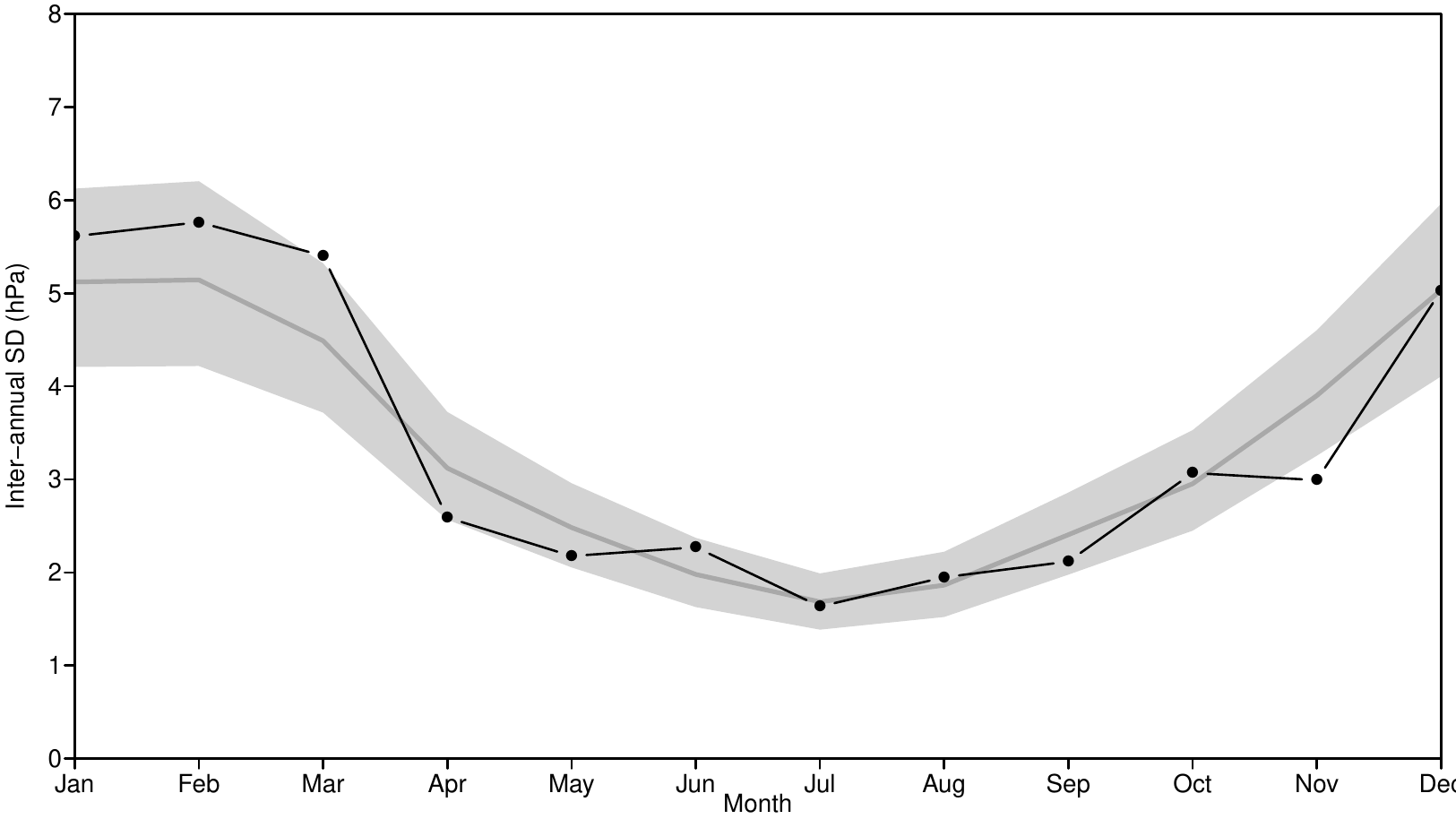}
              \includegraphics[scale=0.5]{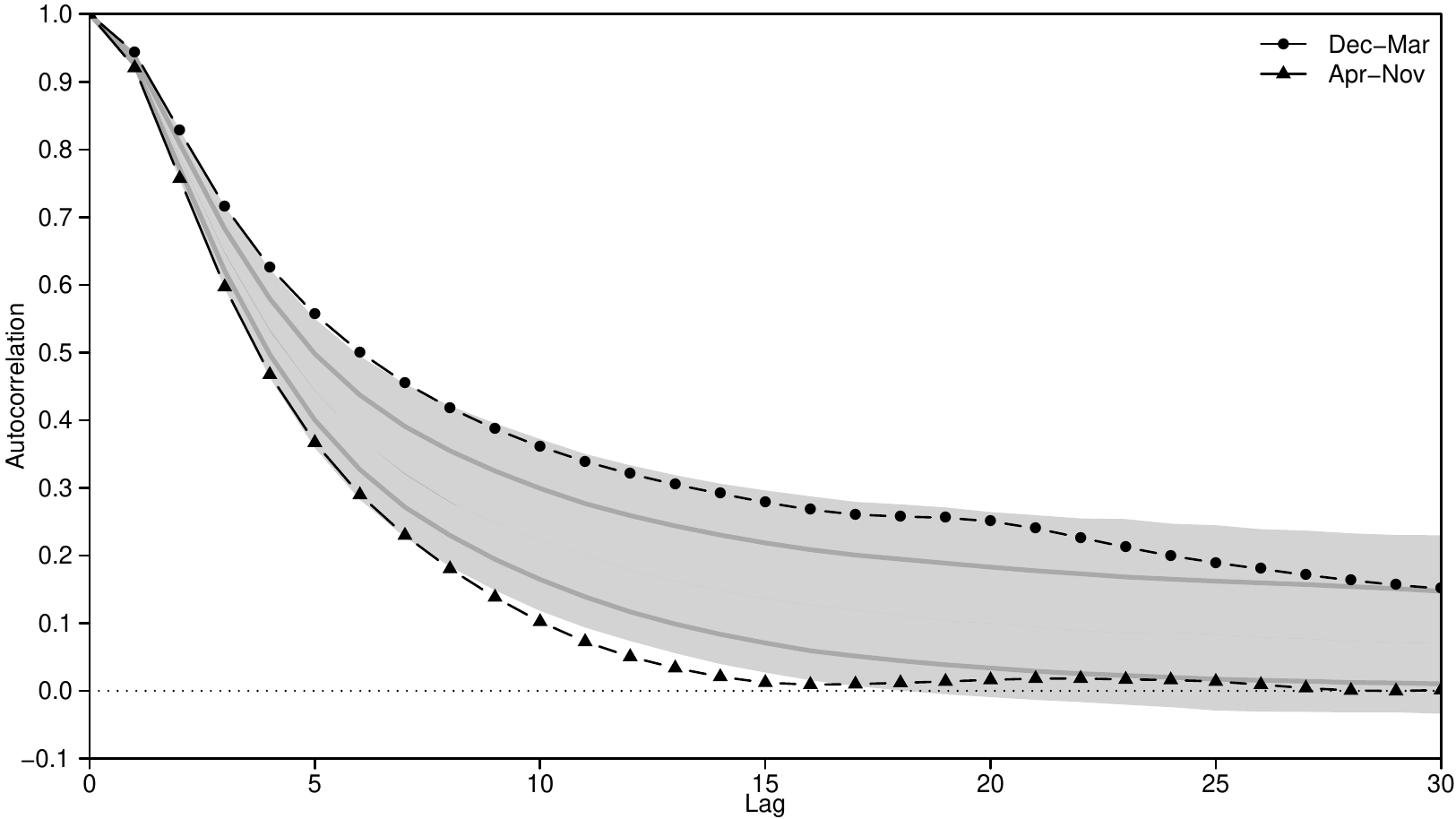}}
  \caption{As in Figure~\ref{fig:postpred} but for the model with coupled period reduced by two weeks.}
  \label{fig:postpred2}
\end{figure}

Figure~\ref{fig:postpred} shows the results of the posterior predictive checks on the initial model with $K = 2$, $P=5$ and 180 day forced period (Nov-Dec-Jan-Feb-Mar-Apr).
Posterior predictive simulations were started from 1968 to give the model time to learn the systematic components (mean, annual cycle, TVAR coefficients).
The model appears to capture the inter-annual variance in Figure~\ref{fig:postpred}(a).
Figure~1(c) of the main text showed a distinct difference in the autocorrelation structure between Dec-Mar and Apr-Nov, therefore we focus on the autocorrelation functions of these two periods.
Figure~\ref{fig:postpred}(b) shows that the model with $P=5$ is unable to reproduce the observed autocorrelation structure during Apr-Nov.
The simulated autocorrelation is more like that of Dec-Mar.
Examining the autocorrelation functions by month revealed that the shift was due to the simulated autocorrelation during April being more like Dec-Mar than May-Nov.
This suggests that the length of the forced is too long.
Figure~\ref{fig:postpred2} shows that reducing the end of the forced period by two weeks to mid-March increases the separation between the simulated  autocorrelation functions and improves the simulation of the inter-annual variance.
However, the model is still unable to properly capture the autocorrelation structure in Apr-Nov.
Fitting additional models with autoregressive components of varying order ($P=3,4,5,6,7$) revealed that the model with $P=6$ is able to capture the observed autocorrelation structure.

\section{Persistence forecasts}

\begin{figure}[t]
  \centerline{\includegraphics[scale=0.5]{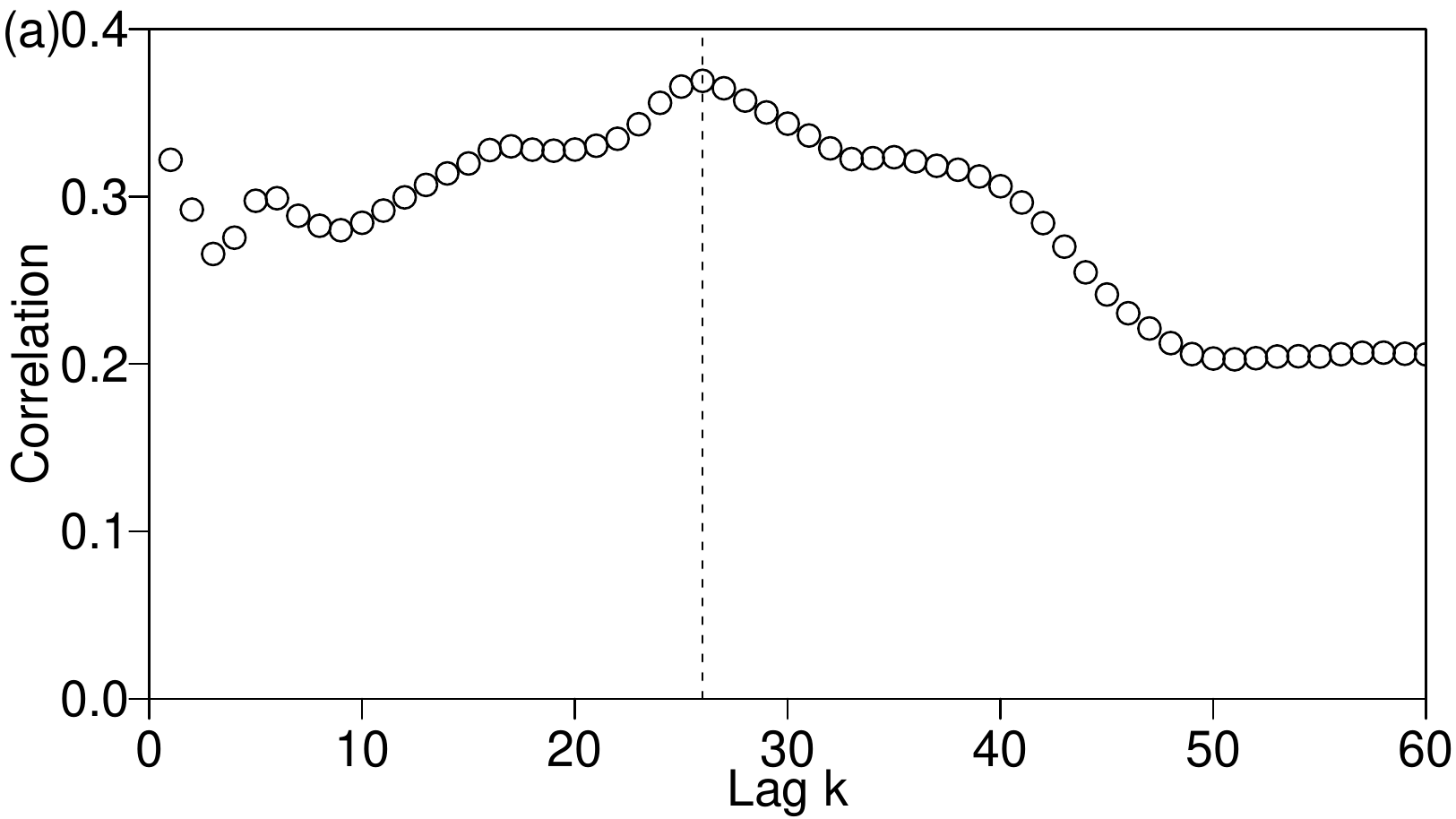}
              \includegraphics[scale=0.5]{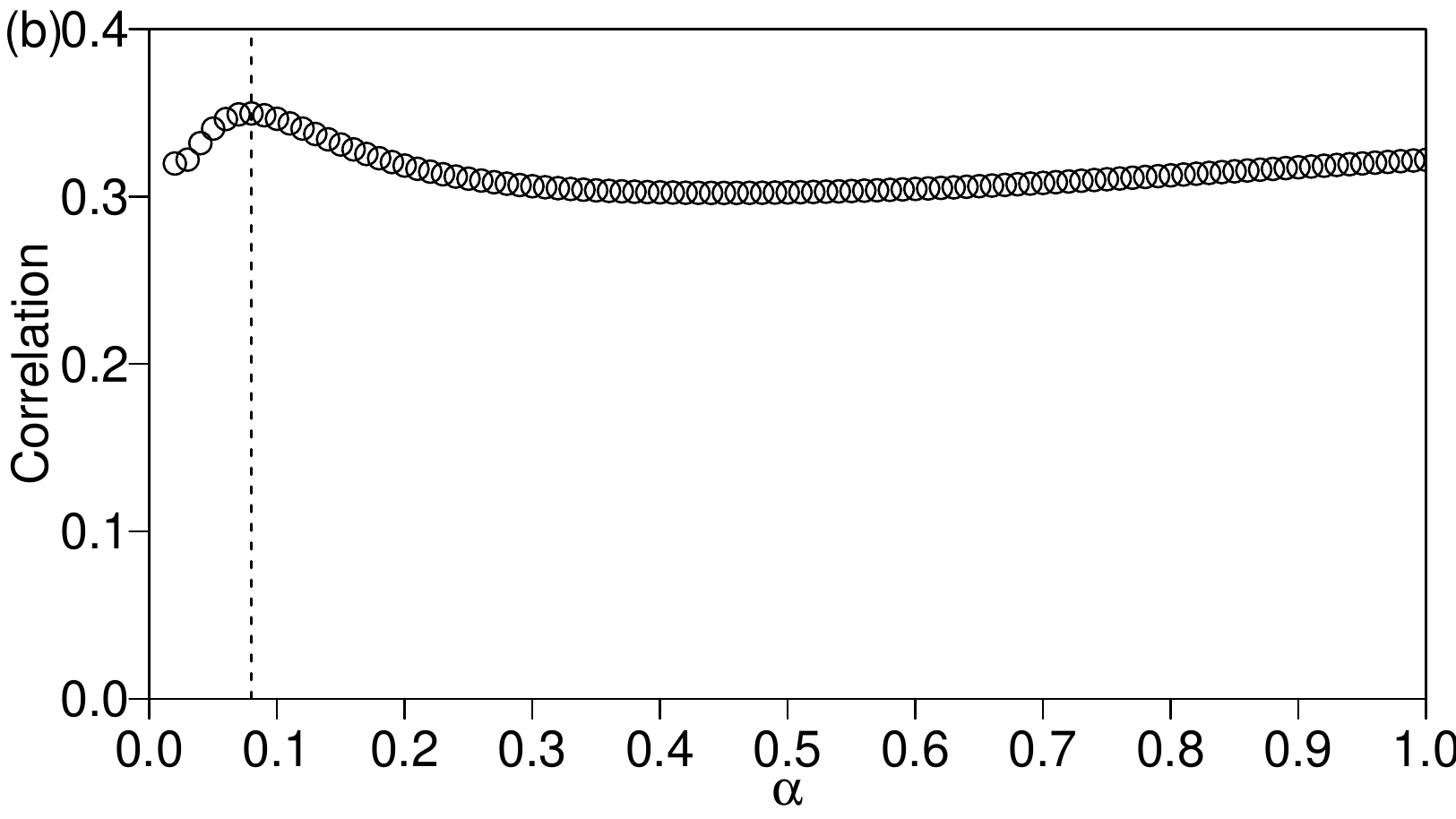}}
  \caption{Optimal persistence forecasts. Comparison of 
           (a) different lags $k$ for the linearly weighted forecast, and
           (b) different parameters $\alpha$ for the exponentially weighted forecast.
           The mean, annual cycle and semi-annual cycle were estimated by least squares and removed from the observations before fitting.}
  \label{fig:persistence}
\end{figure}

The linearly weighted persistence forecast $\hat{Y}_L$ at time $t$ is defined as the mean of the previous $K$ observations
\begin{equation*}
  \hat{Y}_{L t} = \frac{1}{K} \sum_{k = 1}^K Y_{t-k}
\end{equation*}
the optimal value of $K$ was selected as the value that maximized the correlation between the observed winter (DJF) means and the forecasts in Figure~\ref{fig:persistence}(a).

The exponentially weighted persistence forecast $\hat{Y}_E$ at time $t$ is defined by the recursion
\begin{align*}
  \hat{Y}_{E 1} & = Y_1 &  \\
  \hat{Y}_{E t} & = \alpha Y_t + (1 - \alpha) \hat{Y}_{E,t-1} & t > 1
\end{align*}
the optimal value of $\alpha$ was selected as the value that maximized the correlation between the observed winter (DJF) means and the forecasts in Figure~\ref{fig:persistence}(b).

It should be noted that the correlations achieved by the persistence forecasts are probably over-estimates, since no cross-validation design was used.

\bibliographystyle{abbrvnat}
\bibliography{library}